\RequirePackage{fix-cm}
\documentclass[smallcondensed]{svjour3}     
\smartqed  
\usepackage[pdftex]{graphicx}
\usepackage{natbib}

\begin{document}

\title{Development and validation of a transition model based on a mechanical approximation.
}

\titlerunning{Transition V-model}        

\author{Rui Vizinho         \and
        Jos\'{e} P\'{a}scoa \and
	Miguel Silvestre
}


\institute{Rui Vizinho \at
	      Dep. of Electromechanical Engineering
              University of Beira Interior, \\
	      R. Marqu\^{e}s D'Avila e Bolama, 6201-001 Covilh\~{a}\\ \\
              Tel.: +351 96 6940582\\
              Fax: \\
              \email{ruivizinhodeoliveira@sapo.pt}           
           \and
           Jos\'{e} P\'{a}scoa \at
              Dep. of Electromechanical Engineering
              University of Beira Interior, \\
	      R. Marqu\^{e}s D'Avila e Bolama, 6201-001 Covilh\~{a}\\ \\
	   \and
           Miguel Silvestre \at
              Dep. of Aerospace Sciences
              University of Beira Interior, \\
	      R. Marqu\^{e}s D'Avila e Bolama, 6201-001 Covilh\~{a}\\ \\
}

\date{Received: date / Accepted: date}

\maketitle

\begin{abstract}
A new 3D transition turbulence model, more accurate and faster than an empirical transition model, is proposed.
The model is based on the calculation of the pre-transitional $\overline{u^{'}v^{'}}$ due to mean flow shear.
The present transition model is fully described and verified against eight benchmark test cases.
Computations are performed for the ERCOFTAC flat-plate T3A, T3C and T3L test cases.
Further, the model is validated for bypass, cross-flow and separation induced transition and compared with empirical transition models.
The model presents very good results for bypass transition under zero-pressure gradient and with pressure gradient flow conditions.
Also the model is able to correctly predict separation induced transition.
However, for very low speed and low free-stream turbulence intensity the model delays separation induced transition onset.
The model also shows very good results for transition under complex cross-flow conditions in three-dimensional geometries.
The 3D tested case was the 6:1 prolate-spheroid under three flow conditions.
\keywords{Transition model \and ERCOFTAC \and V-model \and Spalart-Allmaras \and OpenFoam \and Three-dimensional}
\end{abstract}

\section{Introduction}
\label{intro}
This work is the result of a continuous research effort on turbulence transition models development, \cite{Vizinho2013_SAE}, \cite{Vizinho2013_EUCASS} and \cite{Vizinho2012_MEFTE}.
Transition modeling is still not widely used in industrial flow computations, mainly due to several limitations of present day models.
This seems to be the case since the available transition models tend to be complex.
Thus, computation time is largely increased by the usage of a transition tool just to calculate an often small fraction of the relevant flow.
It goes without saying that it is of vital importance to know where transition occurs.
However, for some engineering problems this transition onset is overlooked and fully turbulent models are applied instead, \cite{Pascoa2006} and \cite{Pascoa2010}.
Some of these, specifically low-Reynolds turbulence models, are sometimes used to predict transition onset.
The few first serious efforts to evaluate transition modeling capabilities of used turbulence models were presented by \cite{Savill1993a}.
Although some of these models have an apparent transition behavior, \cite{Rumsey2007} presented a work showing that this behavior is mere coincidence and can lead to design mistakes. 
It is then of great interest to have simple robust transition onset prediction tools.

The use and general interest in turbulence transition modeling and control in industrial engineering is steadily increasing, as seen in the works of \cite{B.Aupoix2011}, \cite{Langtry2006}, \cite{Xisto2012}, \cite{Xisto2013} and \cite{Abdollahzadeh2014}.
This also applies for the MAAT project presented by \cite{Ilieva2012}, \cite{Trancossi2012} and \cite{Ilieva2014}, a revolutionary concept of transportation.
Due to its high altitude flight conditions, low turbulence intensities are predominant.
The delay of transition to turbulence is inversely proportional to the free-stream turbulence intensity.
This implies that for the MAAT project vehicles, feeders and cruiser, large regions of laminar flow will occur.
Thus, elements of the feeders and cruiser vehicles, such as propulsive systems, will be exposed to late transition and consequently considerable sized laminar flow regions.
Possession of a turbulence model that can accurately predict transition onset, as well as correctly
compute the transition length, has become a must amongst CFD engineers who deal with transitional
flows in a daily basis. 

A common approach used to tackle these flows was the definition of laminar and turbulent regions on a priori knowledge of the transition zones.
Though functional, this approach could only compute turbulent flow over the extension previously defined by the user.
Also, transition length was not taken into consideration in this method.
This methodology was impractical in prototype cases, since there is no initial information regarding transition onset.

Then, through the simultaneous work of \cite{Smith1956} and \cite{Ingen1956}, the $e^n$ method was developed.
Considered by some to be the state of the art in prediction of transition
onset, it is a method based on the linear local stability theory.
According to \cite{Ingen2008}, ``The linear stability theory considers a given laminar main flow upon which small disturbances are superimposed''. 
The operation protocol of this method begins with the
resolution of the laminar boundary layer velocity profiles.
This is then used to compute the growth rate
of the superimposed linear instabilities within the flow by solving the linear stability equations.
After completion of the latter, an integration of the obtained growth rates throughout
the streamlines is performed. The resulting value will represent the amplification factor, $n$, of the instabilities.
The disturbance amplitude ratio is then given by $e^{n}$. When this variable reaches a threshold, transition onset is assumed.
Typical transition threshold values for $e^{n}$ are obtained with $n$ values between 7 to 9.

Alternatives for transition prediction were proposed, such as in the experimental work of \cite{Skramstad1948} that resulted in one of the first empirical graphic correlations for transition onset.
Later, the works of \cite{Abu-Ghannam1980}, \cite{Mayle1991} and others produced various empirical correlations for transition onset.
Based on these empirical correlations, transition to turbulence models were formulated such as the ones created by \cite{Cho1993}, \cite{Suzen2000} and \cite{Steelant2001}. 
Typically these first empirically correlated transition models were non-locally formulated, since they depended on integration of the boundary layer velocity profiles. 
The latter was performed so as to obtain one of the parameters used by the empirical correlations.
Most commonly this would be the local momentum thickness Reynolds number.
Recently, according to the work of \cite{Menter2002} and \cite{Langtry2006phd}, locally formulated empirical correlation transition to turbulence 
models have been developed.
The previous cited work presented a transition model able to compute transition onset as well as transition length.
The improvement over the previous empirical transition models is that in this formulation the variables are locally defined. 
In \cite{Langtry2006phd} a transition to turbulence closure is coupled with the SST-k-$\omega$ turbulence model of \cite{Menter1994}.
Although, according to the former, it is possible to use these transition components of the model with other turbulence models.
The model formulation can be resumed down to two transport equations for the transition part of the model. 
One for $\gamma$, or intermittency, and another for Re$_{\theta}$, the transition momentum thickness Reynolds number. 
A threshold for momentum thickness Reynolds number is first computed in the free-stream.
This is done according to the employed empirical correlation.
This value is then diffused into the boundary layer.
The momentum thickness Reynolds number is computed locally with an algebraic relation using a vorticity Reynolds number as presented in the works of \cite{VanDriest1963} and \cite{Menter2002}.
When the latter computed value reaches the diffused transition threshold momentum thickness Reynolds number, transition onset is assumed.
The production term in the intermittency transport equation is activated.
This in turn is coupled to the selected turbulence model.
Its production and destruction terms are multiplied by the effective intermittency allowing control of transition onset.
The main problem with the work of \cite{Langtry2006phd} was the lacking of two key components of the model, due to proprietary reasons.
However and thanks to the work of \cite{Suluksna2008}, \cite{Suluksna2009} and \cite{PaulMalan2009} these lacking terms were approximated by alternative functions.
Nonetheless, later in \cite{Langtry2009}, the original authors of the $\gamma-R_{e\theta}$ transition model published all of the lacking terms.
This transition model has a very promising future as an engineering tool.
Also, the work of \cite{Durbin20121} presents a transition prediction tool based on turbulence intermittency that is not empirically correlated.
It is an intermittency transport equation that does not rely on external data input, such as empirical correlations.
Similarly to the latter described model, this equation can be coupled to any turbulence model making it a versatile tool for transition prediction.

Turbulent flow has always been distinguishable from its laminar state.
Transition can be identified as an increase in skin-friction and velocity profile departure from a Blasius distribution.
There use to be a certainty about the fact that only turbulent flow had velocity fluctuations.
Laminar flow was considered to develop in a stationary fashion.
This all changed with the work of \cite{Dryden1937}.
Considered to be the first attempt to successfully measure velocity fluctuations in the laminar flow region.
It was concluded by \cite{Dryden1937}, that it is impossible to perceive a distinction between laminar and turbulent flow solely based upon velocity fluctuation measurements.
These fluctuations in the laminar flow region were attributed to the presence of a turbulent free-stream.
Later on, the experimental work of \cite{Skramstad1948} also confirmed the presence of laminar fluctuations just before turbulence transition onset.
In the latter experiment, free-stream turbulence intensity was reduced as low as possible.
Contrary to the previously recorded laminar fluctuation values from \cite{Dryden1937}, it was expected that in this particular case, the velocity fluctuations would be greatly reduced.
Since these were related to the free-stream turbulence intensity, the results should bear almost no trace of laminar fluctuations.
This was confirmed for the leading edge regions of the laminar flat-plate flow.
Though, the progressive measurement of fluctuations along the flow direction began to detect weak oscillations in the laminar region.
These increased towards the transition up to the bursting of turbulent spots.
To the best of our knowledge, and according to the award winning paper of \cite{Mayle1997}, the work of \cite{Lin1957} was the first to analytically evaluate the effects of laminar fluctuations over laminar velocity profiles.
This study confirmed the possibility of having velocity fluctuations in a laminar flow, even when maintaining a Blasius velocity profile distribution.
This fact was also documented by \cite{Dyban1976}, \cite{Sohn1991} and \cite{Zhou1995}.
In the work of \cite{Mayle1997} the LKE, or, Laminar-Kinetic-Energy theory was proposed.
Following this development, a new tool for transition modeling was available.
The first transition models based on LKE were developed by \cite{LARDEAU2004} and \cite{Walters2004}.
Others followed this trend, such as \cite{Vlahostergios2009}.
In the publication of \cite{LARDEAU2004}, it is stated that, before the increase of skin-friction due to transition, a growth of fluctuation intensity persists in the upper to medium regions of the laminar boundary layer.
Although these oscillations have their origin in the free-stream turbulence, they do not have the same known fully turbulent ratio of $\frac{\overline{-u^{'}v^{'}}} {k}\approx0.3$.
Instead they present far lower values than the latter.
As discussed in \cite{Mayle1997}, these fluctuations do not belong to a normal turbulent regime.
These are then the laminar kinetic energy fluctuations that appear in the upper medium regions of the boundary layer.
These streamwise fluctuations are Klebanoff modes identified by \cite{Klebanoff1971}.
The production of these fluctuations have particular characteristics.
The boundary layer selectiveness of free-stream turbulent eddy scales filters the broad spectrum free-stream turbulence.
This has been identified as ``shear-sheltering`` and was first described by \cite{R.1996}.
The work of \cite{Jacobs1998}, demonstrates this effect using both the discrete and continuous modes of the Orr-Sommerfeld equation.
The continuous modes are eliminated from the main boundary layer region, being confined to the upper reaches of the latter.
It was also concluded that the boundary layer penetration depth by the continuous modes is inversely proportional to their frequency.
This enforces the concept that only low frequency disturbances are amplified by shear in the pre-transitional laminar boundary layer region.
In the experimental work of \cite{Volino1994}, spectra of fluctuating streamwise velocity $u^{'}$, wall-normal velocity $v^{'}$ and turbulent shear stresses $-u^{'}v^{'}$ were recorded.   
It was found in that work and also in \cite{Volino1997b}, noticeable values of $-u^{'}v^{'}$ in the pre-transitional boundary layer.
These were correlated with peak values of low-frequency wall-normal velocity fluctuations $v^{'}$  of the free-stream turbulence.
As previously mentioned these $-u^{'}v^{'}$ values had lower energy and frequency than those found in fully turbulent boundary layer flow.
As discussed in the work of \cite{Leib1999}, \cite{Klebanoff1971}, first named these streamwise velocity fluctuations as ''breathing modes``.
\cite{Taylor1939}, noticed that these oscillations were related with thickening and thinning of the boundary layer.
The production of these streamwise velocity fluctuations $u^{'}$, are believed to be related to the wall-normal velocity oscillations $v^{'}$ through the ''splat-mechanism`` mentioned by \cite{Bradshaw1994}
or by the concept of ''inactive motion`` proposed by \cite{Townsend1961} and \cite{Bradshaw1967}.
As explained by \cite{Volino1998}, a negative $v^{'}$ velocity fluctuation imposed by a turbulent eddy will momentarily compress the boundary layer, shifting higher speed flow against the wall surface.
This results in an increment of $u^{'}$.
As the turbulent eddy is convected by the flow, the imposed compression effect is diminished resulting in a recovery of the boundary layer to its previous state.

The proposed transition model presents some behaviors of the just described processes.
The most noticeable is the prediction of low negative values of $\overline{u^{'}v^{'}}$ in the upper regions of the pre-transitional boundary layer as will be later shown.
Also the model predicts when these values of $\overline{u^{'}v^{'}}$ pierce the laminar-boundary layer.
This is then known as the transition onset.

The main purpose of this work is to present the rational behind the development of a new transition model, henceforward designated as V-model.
The transition V-model is coupled to the Spalart-Allmaras turbulence model and is designated throughout the present work as V-SA.
First, a short description of the three main turbulence transition mechanisms will be presented.
The general V-SA model coupling is disclosed.
Based on the highlighted physics of transition we then proposed the new mechanical model equations.
The description of the used mechanical model approximation is done by first considering some pre-transitional turbulent kinetic energy relations.
Afterwards the mechanical model approximation is presented in detail.
The development of the transport equation for the pre-transitional turbulent kinetic energy of the proposed transition model is presented.
Finally the detailed transition V-model and Spalart-Allmaras turbulence model coupling is described.
There will also be a validation of the model under various geometries and turbulent flow conditions.
Initially the model is validated for zero-pressure-gradient bypass transition using the flat-plate T3A test case.
Then the model is validated for pressure-gradient bypass transition using the flat-plate T3C3 test case.
Thereafter the model is validated for separation induced transition by using the flat-plate T3L test cases.
Finally a validation of transition under cross-flow effects is performed over the three-dimensional 6:1 prolate-spheroid geometry.

\section{Transition mechanisms}
\label{sec:1}
\begin{figure}
	\centering
			\includegraphics[width=1\textwidth]{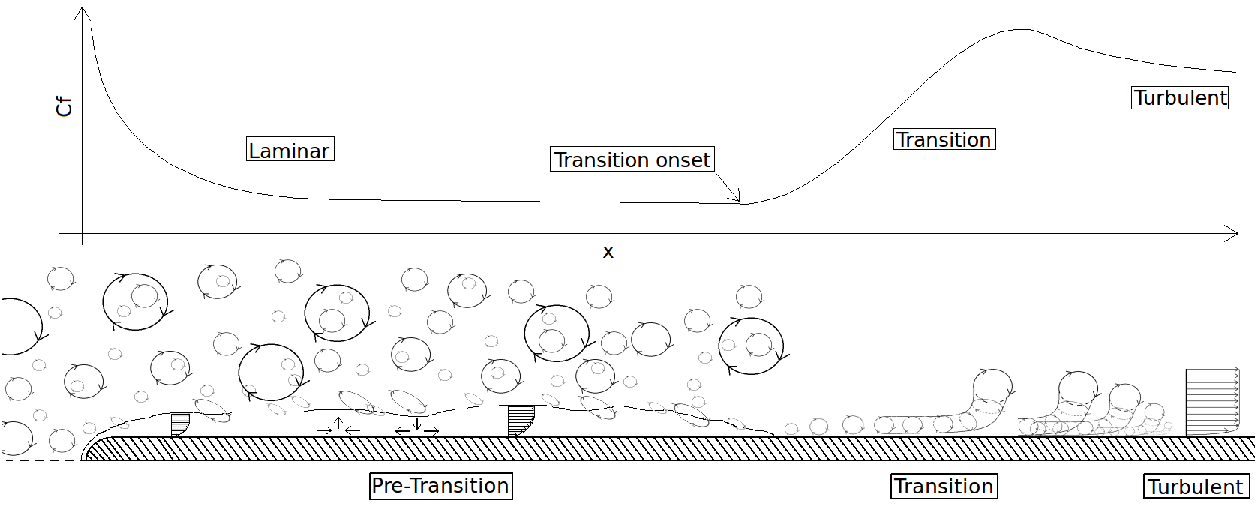}
	
	\caption{General overview of laminar boundary layer transition to turbulence.
Some of the concepts of late transition depicted here were observed in the work of \cite{Liu2011129}.}
\label{fig:Pre_transitional_region_31}
\end{figure}

Generally speaking we may define diverse transition mechanisms, according to the physics of the flow.
In this work we develop a model that focuses on the pre-transitional region depicted in Fig.\ref{fig:Pre_transitional_region_31}. 

\subsection{Natural transition} 
\label{sec:2}
For flow with free-stream turbulence intensity, FSTI$\leq1\%$, natural transition is generally observed when the developing laminar flow reaches a critical Reynolds number value. 
After this critical stage of flow development, viscous Tollmien-Schlichting instability waves begin to slowly grow, from small linear perturbations to non-linear disturbance waves. 
Having reached this point, the critical location is usually defined where the first instability starts to grow, and the transition point where the first turbulent spot appears in the transition process.
The non-linear waves create three dimensional disturbances and, by means of inviscid mechanisms, spots of turbulence begin to appear randomly inside the laminar boundary layer. 
Some of these spots grow in size and downstream from where transition first began to develop, a process of fusion between these spots gives light to the fully turbulent boundary layer. 

\subsection{Bypass transition}
\label{sec:3}
Bypass transition may be sub-critical, it may occur upstream of the previously defined critical location.
Bypass transition normally occurs due to two sources of disturbances, surface roughness and high FSTI, which translates to FSTI$\geq1\%$. 
The bypass transition mechanism can be shortly described as a natural transition process without the Tollmien-Schlichting wave development phase. 
Therefore, laminar flow under bypass transition will turn from laminar to the turbulent spot surge phase immediately.
The subsequent development is similar to the natural transition.

\subsection{Separation induced transition}
\label{sec:4}
Separation of flow is generally observable in flow development over surfaces under adverse pressure-gradient conditions. 
Due to flow separation a vortex bubble is formed and flow may, or may not, reattach by closing the bubble. 
The reattachment process is dependent on the increase of flow mixture capacity originated by turbulence.
Laminar flow transition to turbulence occurs in the separated shear layer.
The shear layer gains momentum from the free-stream and reattaches to the wall. 
Separation bubble length is inversely proportional to FSTI. 

\section{Transition model coupling} 
\label{sec:5}
\begin{figure}
	\centering
			\includegraphics[width=0.6\textwidth]{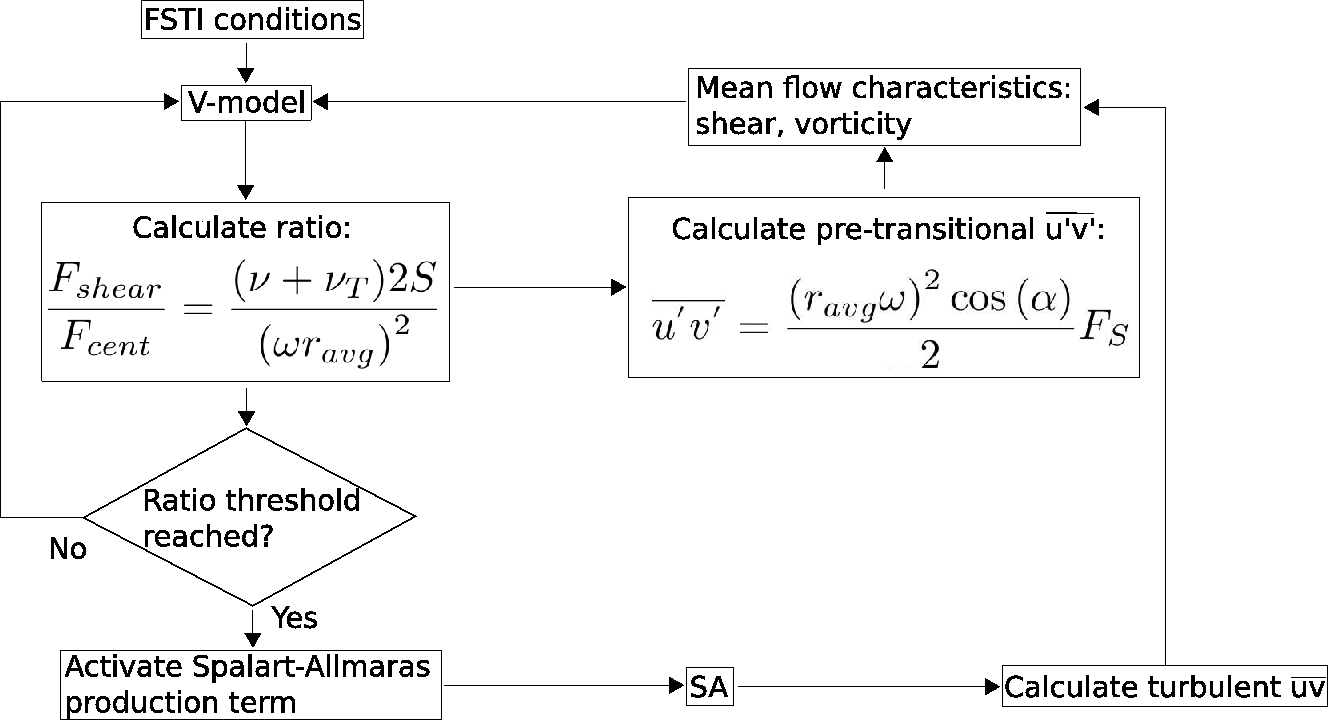}
	
	\caption{V-SA model architecture, by coupling the transition V-model to the Spalart-Allmaras turbulence model.}
\label{fig:Esquematico_V_SA}
\end{figure}

The V-model is not able to compute turbulence.
Instead it determines the transition threshold region.
For this reason and as previously mentioned, the V-model transition closure was coupled to the Spalart-Allmaras turbulence model.
Transition onset prediction is performed by computing the viscosity induced by the predicted pre-transitional $\overline{u^{'}v^{'}}$ values described throughout this work.
The modus-operandi of the V-SA model is depicted in Fig.\ref{fig:Esquematico_V_SA}.

\section{Mechanical model approximation rational}
\label{sec:6}
The rational behind the development of the transition V-model is herein presented, including the flow physics on which it is supported.
Before the mechanical approximation disclosure, some considerations need to be taken into account and explained.
It is here assumed that pre-transitional turbulence is isotropic in a strain-less free-stream, in the sense that $k_{x}=k_{y}=k_{z}=k_{p}$.
This is in agreement with the flow physics of transition such as presented in the work of \cite{Mayle1997}.
However, under the effect of flow shear the model will predict small pre-transitional negative values of $\overline{u^{'}v^{'}}$ related to non-isotropic turbulence conditions.
A bi-dimensional analysis is here described.
Nevertheless the V-model transition closure is applicable to three-dimensional cases as will be later presented.
This is the case since the main three orthogonal shear deformation planes are all accounted for in the computation of the local shear magnitude.
As such, a global effect of three-dimensionality is taken into account in the computation of the pre-transitional $\overline{u^{'}v^{'}}$.  

\subsection{Pre-transitional turbulent kinetic energy considerations}
\label{sec:7}
\begin{figure}
	\centering
			\includegraphics[width=0.55\textwidth]{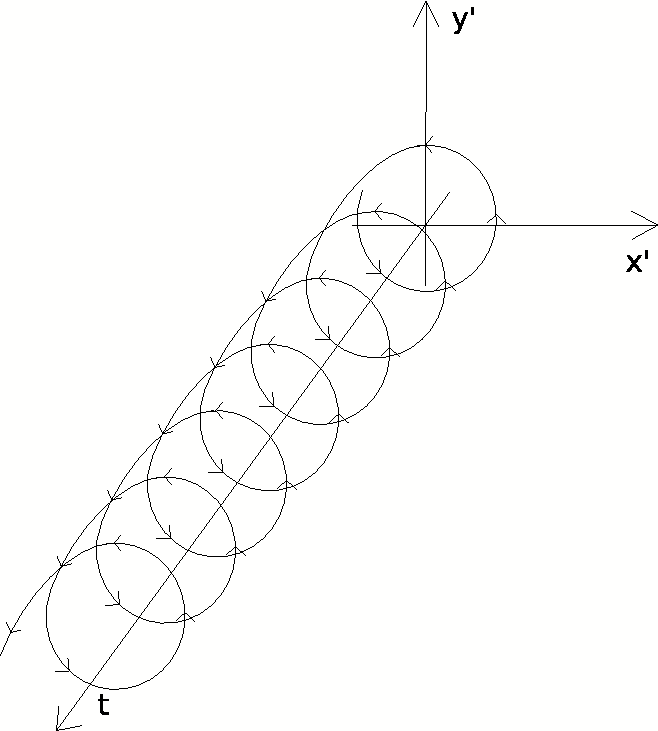}
	
	\caption{Particle trajectory describing a circular path through time.}
\label{fig:Esquematico_trajectoria_xyt}
\end{figure}

Pre-transition velocity fluctuations wave forms, for a specific frequency, have seldom a regular shape.
Although this is true for most cases, the modeling of the pre-transition region, see works such as \cite{Jacobs1998}, requires both discrete and continuum modes.
Citing \cite{Jacobs1998}, ''The eigensolutions to the Orr-Sommerfeld equation in an unbounded domain are classified into two spectra: the first is a finite set of discrete modes; the second is an infinite continuum of modes.
The latter are weakly damped and are irrelevant to classical linear stability analysis.
Unstable modes are only members of the discrete spectrum.``
As previously stated, the modes used in classical linear stability analysis are the discrete spectrum components. 
However, the present transition model will attempt to model the effects on the continuum spectrum of modes.
Citing \cite{Jacobs1998}, ''The eigenfunction of the discrete modes decays exponentially with distance above the boundary layer.
The eigenfunction of the continuous modes is sinusoidal in that region.''
Therefore and in order to simplify the following exposure a sinusoidal wave shape was considered to model the free-stream pre-transition continuum spectrum.
It is here assumed that sinusoidal wave forms represent the time evolution of velocity fluctuations due to pre-transitional turbulence.
Admitting that a particle is stuck inside one of these special pre-transitional turbulent vortices, its movement follows that of the vortex.
Considering then a cross sectional plane of the bi-dimensional vortex, the equations of motion for the particle imprisoned in the small pre-transitional vortex can be obtained.
The equations of motion for $x^{'}$ and $y^{'}$ are then defined by (\ref{equacao_x}) and (\ref{equacao_y}) respectively.
These were obtained considering as a frame of reference the center of the pre-transitional vortex itself.  
\begin{equation}
x^{'}
=
-r_{avg} \cos\left(\omega t+\frac{\alpha}{2} \right),
\label{equacao_x}
\end{equation}
\begin{equation}
y^{'}
=
-r_{avg} \cos\left(\omega t-\frac{\alpha}{2} \right).
\label{equacao_y}
\end{equation}
The time derivative of the latter will introduce the equations of the velocity fluctuations $u^{'}$ and $v^{'}$ represented by (\ref{equacao_u}) and (\ref{equacao_v}).
\begin{equation}
u^{'}
=
r_{avg} \omega \sin\left(\omega t+\frac{\alpha}{2} \right),
\label{equacao_u}
\end{equation}
\begin{equation}
v^{'}
=
r_{avg} \omega \sin\left(\omega t-\frac{\alpha}{2} \right).
\label{equacao_v}
\end{equation}
As already mentioned, it should be noted that these last four laws of motion are deduced assuming a frame of reference of the vortex itself.
From the presented equations (\ref{equacao_x}-\ref{equacao_v}), consideration of the $\alpha$ values must be taken.
It can be seen that for $\alpha=\frac{\pi}{2}$ these laws of motion describe a circular motion as shown in Fig.\ref{fig:Esquematico_trajectoria_xyt}. 
This motion can be interpreted as a circular non-deformed pre-transitional vortex.
Nonetheless, for $\alpha=\frac{\pi}{4}$ and $\alpha=\frac{3\pi}{4}$ the described motion will be elliptical.
The limiting cases are obtained for $\alpha=0$ and $\alpha=\pi$, where although the motion is periodical, it is also linear.

\begin{figure}
	\centering
			\includegraphics[width=0.55\textwidth]{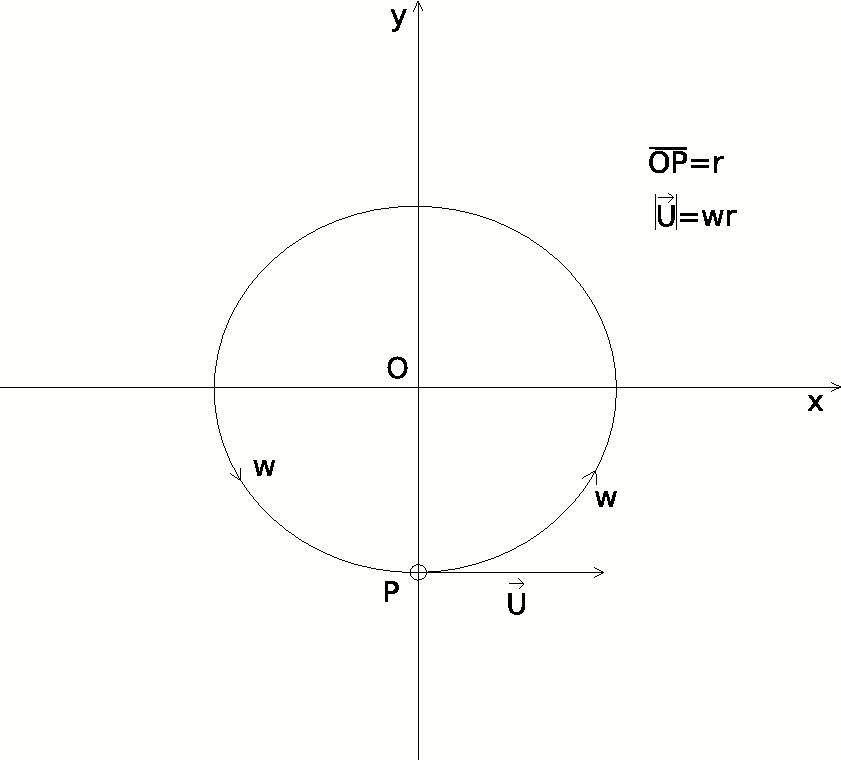}
	
	\caption{The pre-transitional turbulent vortices can be generated at the interface between the free-stream turbulence and the pre-transitional laminar boundary layer due to ''inactive motion``, \cite{Bradshaw1994}.
It is here presented a cut-section of a circular pre-transitional vortex.}
\label{fig:Esquematico_energia_vortice}
\end{figure}

The pre-transitional vortex will have its turbulent kinetic energy.
This can be related to a more simple definition of kinetic energy.
In order to perform the analogy, we take another look at the clinging particle under the effect of the circular pre-transitional vortex.
This circular motion analogy is considered under a bi-dimensional plane coincident with the xy Cartesian frame of reference positioned within the vortex center.
Considering Fig.\ref{fig:Esquematico_energia_vortice}, the animated particle in point P will have a certain amount of specific kinetic energy defined as, $e_{c_{P}}=U^{2}/2$.

This same particle will have a mean turbulent kinetic energy since it follows the pre-transitional vortex rotational motion.
The latter is defined as the sum of the mean turbulent kinetic energy along xx's axis and the mean turbulent kinetic energy along yy's axis.
Assuming a circular vortex, and taking this into consideration, through (\ref{equacao_conservacao_energia}), a relation can be obtained in order to characterize the value of rotational velocity and radius of the pre-transitional vortex.
The resulting relation is then (\ref{equacao_conservacao_energia_resultado}).

\begin{equation}
k_{x}+k_{y}
=
\frac{1}{2}\overline{u^{'}u^{'}}+\frac{1}{2}\overline{v^{'}v^{'}}=e_{c_{P}}=\frac{\left(\omega_{z} r_{avg}\right)^{2}}{2},
\label{equacao_conservacao_energia}
\end{equation}

\begin{equation}
\sqrt{2\left( k_{x}+k_{y}\right)}
=
\omega_{z} r_{avg}.
\label{equacao_conservacao_energia_resultado}
\end{equation}

In (\ref{equacao_conservacao_energia_resultado}), $\omega_{z}$ represents the vortex rotational speed and $r_{avg}$ is the average pre-transitional vortex radius in the xy cross section plane.

An equivalence of turbulent kinetic energy and kinetic energy was just performed.
Besides the assumption of a circular pre-transitional vortex, up until this point no approximation has been used.
However, in order to close this particular system an approximation must be performed.
Either the average radius of the vortices is approximated or the rotational velocity of the vortices, $\omega_{z}$.
The approach of rotational velocity approximation was chosen.
The used relation was defined on the assumption that, rotational velocity of the pre-transitional turbulent vortices should be proportional to its pre-transitional turbulent kinetic energy.
Also, the rotational velocity of pre-transitional vortices should be inversely proportional to the flow kinematic viscosity.
This is further related to the fact that for a fixed turbulent large scale, the small turbulence sizes, such as the Kolmogorov scales, are reduced with increasing Reynolds number flows.
The latter is also related to an increase of the turbulent vortices rotational velocities due to angular momentum conservation.
Therefore, the lower the fluid kinematic viscosity, the higher the flow Reynolds number and consequently the higher the turbulent vortices rotational velocities.
Besides this, dimensional analysis of the used relation confirms its correct dimensional characteristics.
The selected approximation is then presented in (\ref{equacao_Wxy}).
\begin{equation}
\omega_{z}
=
\frac{k_{x}+k_{y}}{\nu}.
\label{equacao_Wxy}
\end{equation}
Throughout the remaining model exposure $\omega_{z}=\omega$.
Therefore, since $k_{x}+k_{y}=2k_{p}$, the value of $\omega$ is now computed using the relation presented in (\ref{equacao_omega}). 
\begin{equation}
\omega
=
\frac{2k_{p}}{\nu}.
\label{equacao_omega}
\end{equation}
From the latter relation and equation (\ref{equacao_conservacao_energia_resultado}), the value of $r_{avg}$ is calculated in (\ref{equacao_r_avg}).
\begin{equation}
r_{avg}
=
\frac{\sqrt{4k_{p}}}{\omega}.
\label{equacao_r_avg}
\end{equation}
In order to compute terms such as $\overline{u^{'}u^{'}}$ or $\overline{u^{'}v^{'}}$ 
a time average of these fluctuating values must be performed according to equation (\ref{equacao_uu_energia_generica}).

\begin{equation}
\overline{u^{'}u^{'}}
=
\lim_{T\to\infty}\frac{1}{T}\int_0^Tu^{'}u^{'}dt.
\label{equacao_uu_energia_generica}
\end{equation}
  
It was then assumed a sinusoidal function for these fluctuating velocities.
This purports that these fluctuations have a periodicity.
The latter implies that a finite value for $T$ is possible. 
A plausible value for $T$ could be the periodicity of these sinusoidal functions.
The latter assumption results in (\ref{equacao_uu_energia}).

\begin{equation}
\overline{u^{'}u^{'}}
=
\frac{\omega}{2\pi}\int_0^\frac{2\pi}{\omega} u^{'}u^{'}dt.
\label{equacao_uu_energia}
\end{equation}

Using the explicit formulations of $u^{'}$ and $v^{'}$, the calculation of $\overline{u^{'}u^{'}}$ and $\overline{u^{'}v^{'}}$ 
should be done according to (\ref{equacao_uu_energia_explicita}) and (\ref{equacao_uv_energia_explicita}) respectively.

\begin{equation}
\overline{u^{'}u^{'}}
=
\frac{\omega}{2\pi}\int_0^\frac{2\pi}{\omega} \left[r_{avg} \omega \sin\left(\omega t+\frac{\alpha}{2} \right)\right]^{2}  dt,
\label{equacao_uu_energia_explicita}
\end{equation}

\begin{equation}
\overline{u^{'}v^{'}}
=
\frac{\omega}{2\pi}\int_0^\frac{2\pi}{\omega} r_{avg} \omega \sin\left(\omega t+\frac{\alpha}{2} \right) r_{avg} \omega \sin\left(\omega t-\frac{\alpha}{2} \right)  dt.
\label{equacao_uv_energia_explicita}
\end{equation}

The $\overline{u^{'}u^{'}}$ value is directly obtained through (\ref{equacao_uu_energia_explicita_resultado}).
It can be seen that for isotropic pre-transitional turbulence, $k_{x}=k_{y}$, equation (\ref{equacao_uu_energia_explicita_resultado}) is in accordance with the kinetic energy relations of (\ref{equacao_conservacao_energia}) and (\ref{equacao_conservacao_energia_resultado}).

\begin{equation}
\overline{u^{'}u^{'}}
=
\frac{r_{avg}^{2}\omega^{2}}{2}.
\label{equacao_uu_energia_explicita_resultado}
\end{equation}

The $\overline{u^{'}v^{'}}$ value is then calculated using relation (\ref{equacao_uv_energia_explicita_resultado}). 
It should be noted that $r_{avg}$ is the mean radius of the undeformed pre-transitional vortex.
As can be seen in (\ref{equacao_uv_energia_explicita_resultado}), there is a dependence with $\alpha$.
This $\alpha$ represents the phase shift between the two velocity components $u^{'}$ and $v^{'}$.
It also represents the deformation angle of the pre-transitional vortex.
For $\alpha$ equal to $\frac{\pi}{2} rad$ or $90\;\mathring{ }$ the pre-transitional vortex has a circular undeformed shape.
Also, for this value of $\alpha$, according to (\ref{equacao_uv_energia_explicita_resultado}), the Reynolds shear stresses, $\overline{u^{'}v^{'}}$, will be zero for an undeformed pre-transitional vortex.

\begin{equation}
\overline{u^{'}v^{'}}
=
\frac{r_{avg}^{2} \omega^{2} \cos\left(\alpha\right)}{2}.
\label{equacao_uv_energia_explicita_resultado}
\end{equation}

\subsection{Mechanical model for pre-transitional turbulent velocity fluctuation components under mean shear}
\label{sec:8}
The fact that $\overline{u^{'}v^{'}}$ has a trend to present negative values under shear influence is commonly accepted.
The reasoning begins by considering a no-slip wall constrained velocity profile.
Under this scenario, a positive vertical velocity fluctuation away from the wall, such that $v^{'}>0$, will induce a reduction of the flow momentum.
The excited particle tends to maintain its momentum, this is lower than the new surrounding fluid due to the presence of the wall no-slip condition.
This will then imply that $u^{'}<0$, since the relative velocity of the new low momentum fluid is lower than the surroundings.
The reverse case is also applicable, that is for $v^{'}<0$ a value of $u^{'}>0$ will follow.
Therefore $\overline{u^{'}v^{'}}$ has a negative value. 

\begin{figure}
	\centering
			\includegraphics[width=0.55\textwidth]{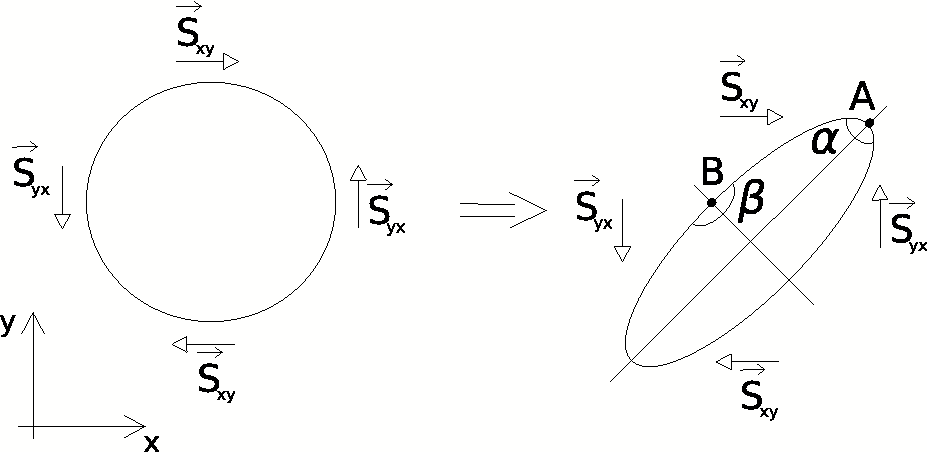}
	
	\caption{The deformation of a large scale vortical structure due to mean flow shear assumes the presented shape.
Also, this is the expected pre-transitional vortex deformation shape under mean flow shear.}
\label{fig:Esquematico_deformacao_circulo}
\end{figure}

\begin{figure}
	\centering
			\includegraphics[width=0.55\textwidth]{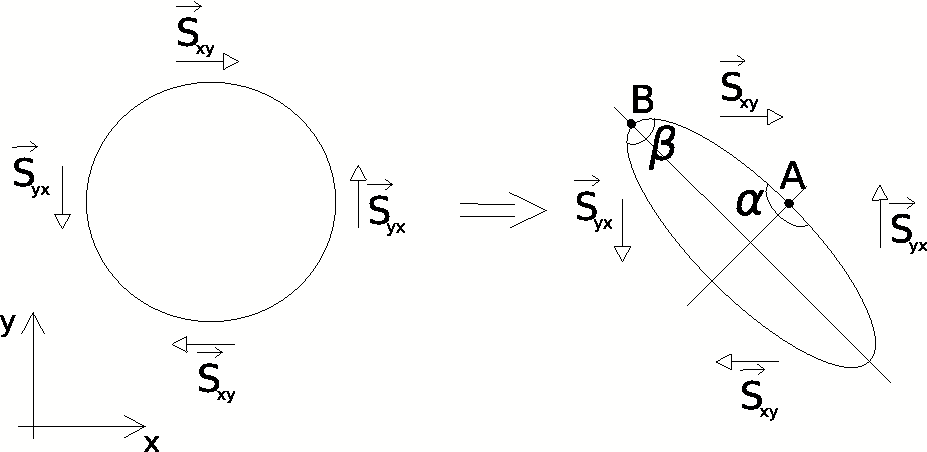}
	
	\caption{The derived relation (\ref{equacao_racio}) for computation of the $\alpha^{'}$ deformation angle is used to calculate the $\alpha$ deformation angle through (\ref{equacao_alfa_linha}).
The former relation predicts an $\alpha$ angle increase with shear, instead of the expected angle decrease as depicted in Fig.\ref{fig:Esquematico_deformacao_circulo}.
As the mean flow shear increases so does the ratio calculated according to (\ref{equacao_racio_abreviatura}).
The predicted pre-transitional vortex deformation under mean flow shear is presented here.}
\label{fig:Esquematico_deformacao_circulo_previsto}
\end{figure}

Consider then the case where an initially circular pre-transitional vortex is deformed by mean shear as in Fig.\ref{fig:Esquematico_deformacao_circulo}.
It must be noted that the presented schematic only demonstrates what is to be expected, not what the mechanical model predicts.
Due to shear deformation, there will be an alteration of path curvature along the vortex surface.
The centrifugal force distribution along the pre-transitional vortex will change accordingly.
These centrifugal forces defined in (\ref{equacao_F_centrifuga}), will act as pseudo non-linear springs of the pre-transitional vortex. 
These will act on the vortex when shear is present.
The effect of the vortex deformation is reflected on the centrifugal forces computation, (\ref{equacao_F_centrifuga}), through the local curvature radius, $r_{local}$.
\begin{equation}
F_{cent}
=
\frac{U^{2}\rho V}{r_{local}}.
\label{equacao_F_centrifuga}
\end{equation}
In order to compute the centrifugal forces, (\ref{equacao_F_centrifuga}), the local radius of curvature, $r_{local}$, needs to be calculated.
The change of the vortex curvature is dependent on the predicted deformation angle, $\alpha$.
As such, a relation between the local radius of curvature and the angle of deformation of the system is required.
The latter relation was developed so as to deliver the requirements in,
\begin{equation}
\alpha
=
\pi \Rightarrow r_{local}
=
\infty,    
\label{equacao_alfa_r_requerimentos_1}
\end{equation}
\begin{equation}
\alpha
=
0 \Rightarrow r_{local}
=
0,
\label{equacao_alfa_r_requerimentos_2}
\end{equation}
\begin{equation}
\alpha
=
\frac{\pi}{2} \Rightarrow r_{local}
=
r_{avg}.
\label{equacao_alfa_r_requerimentos_3}
\end{equation}
The requirement presented in (\ref{equacao_alfa_r_requerimentos_1}) is representative of an absolute flat vortex with orientation depicted in Fig.\ref{fig:Esquematico_deformacao_circulo_previsto}.
The following requirement in (\ref{equacao_alfa_r_requerimentos_2}) represents the scenario of a perfectly flat vortex with orientation depicted in Fig.\ref{fig:Esquematico_deformacao_circulo}. 
The final requirement in (\ref{equacao_alfa_r_requerimentos_3}) represents the undeformed circular vortex.
Also, the calculation of local radius will depend on the average radius of the vortex and its deformation angle, $\alpha$.
The developed relation was then, 
\begin{equation}
r_{local}
=
r_{avg}\frac{\alpha}{\pi-\alpha}.
\label{equacao_r_local}
\end{equation}
This relation enforces the requirements in (\ref{equacao_alfa_r_requerimentos_1}-\ref{equacao_alfa_r_requerimentos_3}).

The shear force acting upon the vortex is defined in (\ref{equacao_F_shear}).
In the latter, S, is mean flow shear magnitude.
The $A$ and $V$ terms in both the equations (\ref{equacao_F_centrifuga}) and (\ref{equacao_F_shear}), represent surface area and volume of the pre-transitional vortex respectively.
These are defined in (\ref{equacao_area}) and (\ref{equacao_volume}).
The $l$ is here considered as the length of the pre-transitional vortex.
A value for this is not required since $l$ will cancel itself in the upcoming model development.

\begin{equation}
F_{shear}
=
(\mu+\mu_{T}) SA,
\label{equacao_F_shear}
\end{equation}

\begin{equation}
A
=
2\pi r_{avg}l,
\label{equacao_area}
\end{equation}

\begin{equation}
V
=
\pi r_{avg}^{2}l.
\label{equacao_volume}
\end{equation}

\begin{figure}
	\centering
			\includegraphics[width=0.7\textwidth]{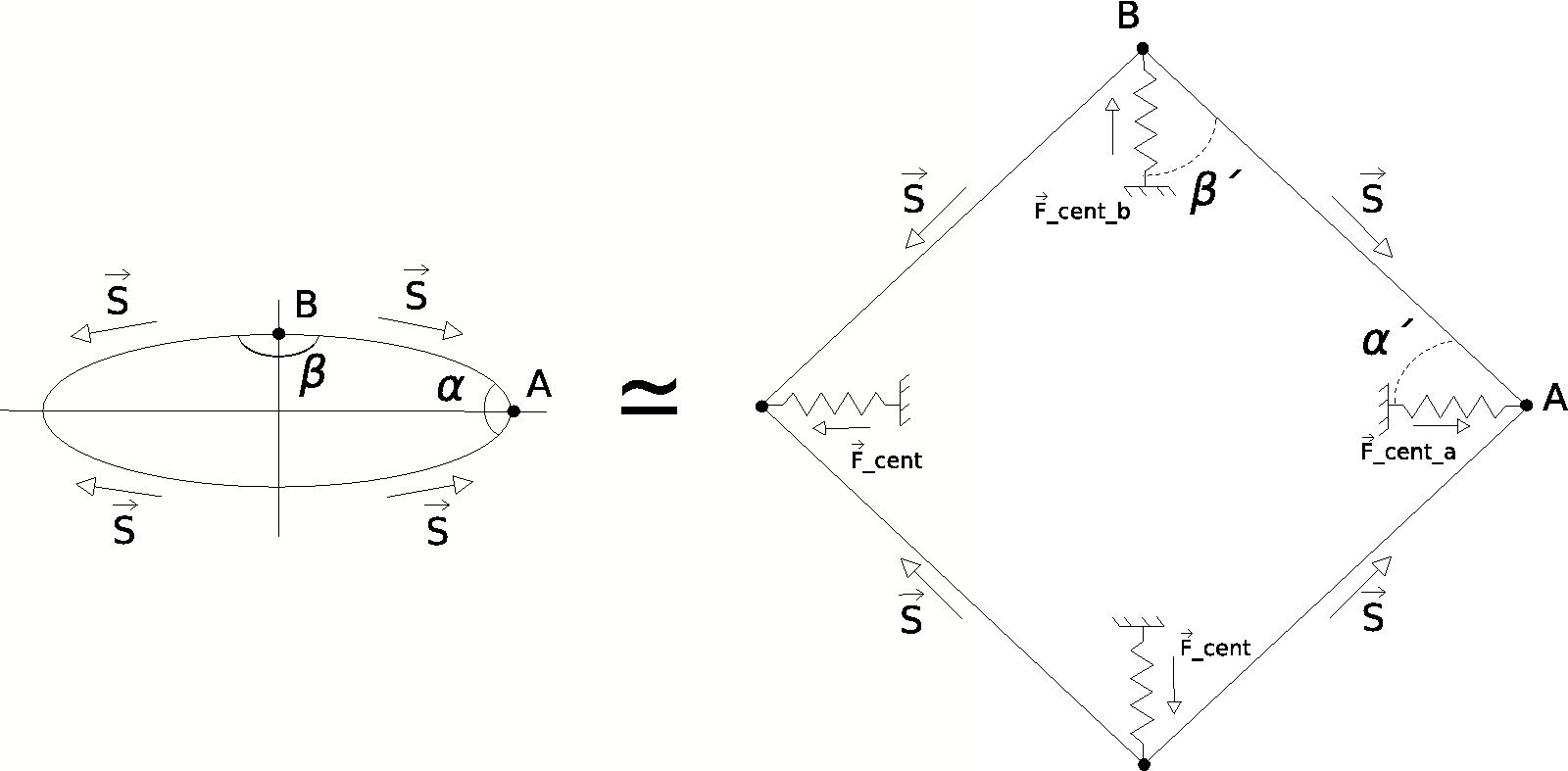}
	
	\caption{The pre-transitional boundary layer oscillations due to shear known as Klebanoff modes were first observed and named ''breathing modes`` by \cite{Klebanoff1971}.
This behavior of the pre-transitional turbulent vortices can be accounted for by the mechanical model approximation.
The depicted mechanical model approximation makes use of a fictitious non-linear spring analogy to describe the internal forces acting on the vortex.}
\label{fig:Esquematico_aproximacao_mecanica}
\end{figure}

\begin{figure}
	\centering
			\includegraphics[width=0.55\textwidth]{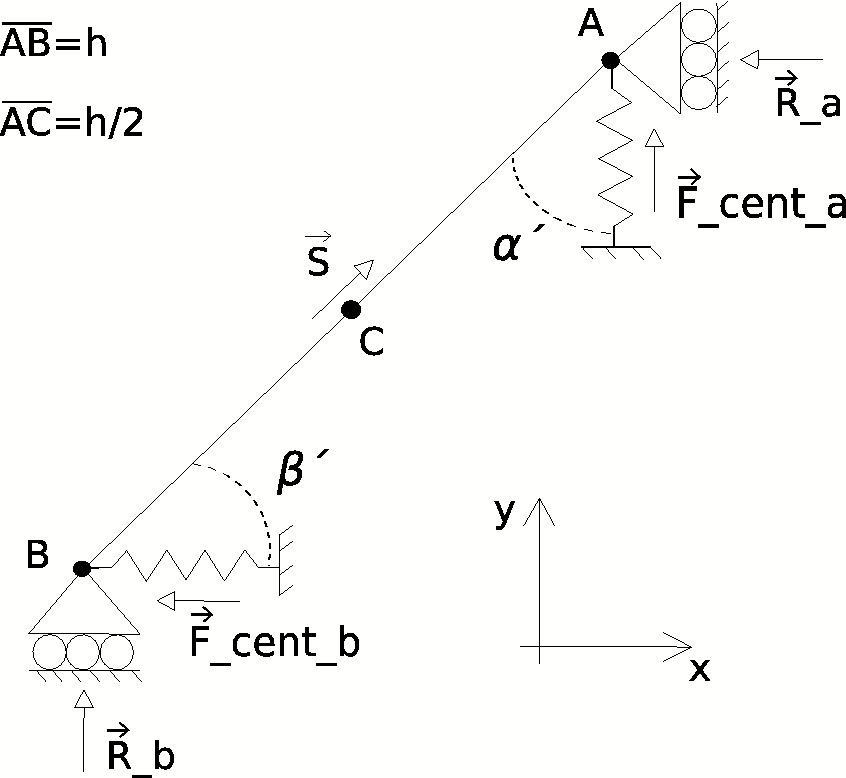}
	
	\caption{The shear force, $\vec{S}$, effect on pre-transitional eddies is approximated by this mechanical model for one quarter of a circular vortex.
The mechanical approximation depicted here is applied to pre-transitional turbulent vortices in the upper zones of the laminar boundary layer.}
\label{fig:Esquematico_sistema_mecanico}
\end{figure}

This non-linear spring feature of the centrifugal forces can be modeled by a relatively simple mechanical system shown in Fig.\ref{fig:Esquematico_aproximacao_mecanica}.
The non-linear spring mechanical behavior is given by the centrifugal force physical characteristics as disclosed in (\ref{equacao_F_centrifuga}).
The $r_{local}$ represents the local curvature radius of the pre-transitional vortex.
In Fig.\ref{fig:Esquematico_aproximacao_mecanica}, the left picture represents the expected statistical mean shape of a pre-transitional vortex under the effect of mean flow shear.
Normally the shear tensor main axes are aligned $45\;\mathring{ }$ with the local flow direction.
The presented $\alpha$ and $\beta$ angles in this left picture of Fig.\ref{fig:Esquematico_aproximacao_mecanica} are representative of the vortex deformation angles located in the directions of the shear stress tensor major axis and minor axis.
Also, this $\alpha$ angle represents the phase shift in the motion equations (\ref{equacao_x}-\ref{equacao_v}).
The right schematic in Fig.\ref{fig:Esquematico_aproximacao_mecanica}, is the approximation of the continuum vortex shown in the left picture by a discrete system composed of four elements.
In this schematic, the $\alpha^{'}$ and $\beta^{'}$ angles are the half values of the original $\alpha$ and $\beta$ angles.
Therefore, the angle relations in this problem are shown in (\ref{equacao_alfa_linha}) and (\ref{equacao_beta_linha}).
\begin{equation}
\alpha
=
2\alpha^{'},
\label{equacao_alfa_linha}
\end{equation}
\begin{equation}
\beta^{'}
=
\frac{\pi}{2}-\alpha^{'}.
\label{equacao_beta_linha}
\end{equation}  
The solution of the mechanical dynamic problem presented in the right schematic of Fig.\ref{fig:Esquematico_aproximacao_mecanica} can be simplified by considering only one quarter of the system.
This is possible due to the double symmetry along the shear stress tensor major axis and minor axis coincident with the axes of the ellipse resulting from the deformed circular pre-transitional vortex.
With this in mind, the final mechanical model approximation is then shown in Fig.\ref{fig:Esquematico_sistema_mecanico}.
The presented orientation is in accordance to Fig.\ref{fig:Esquematico_deformacao_circulo}.

\begin{figure}
	\centering
			\includegraphics[width=0.9\textwidth]{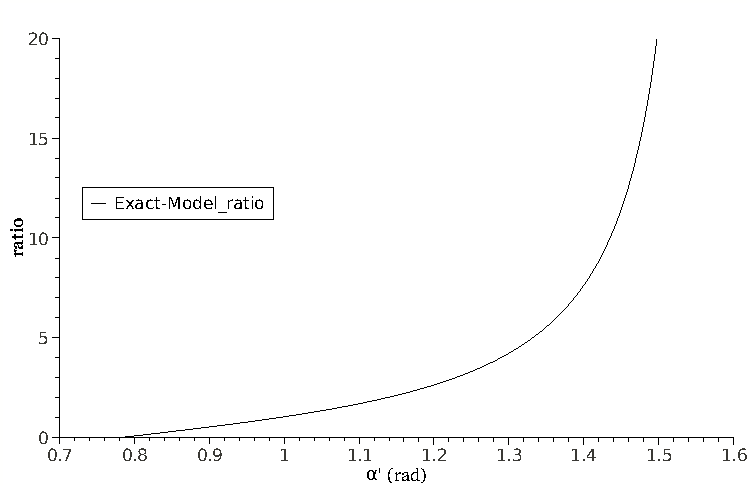}
	
	\caption{Ratio evolution with $\alpha^{'}$ according to the mechanical system solution (\ref{equacao_racio}).}
\label{fig:ratio_evolution}
\end{figure}

The attempt to solve the mechanical system in Fig.\ref{fig:Esquematico_sistema_mecanico} will produce a relation between vortex deformation and mean shear.
In order to solve the mechanical problem presented in Fig.\ref{fig:Esquematico_sistema_mecanico}, classical mechanical system solving procedures must be taken.
First an equilibrium of the system moments in relation to point ''C'' is in order.
The resulting equation is presented in (\ref{equacao_equilibrio_momentos}).
\begin{equation}
\Sigma M_{C}
=
R_{A}\frac{h}{2}cos(\alpha^{'})+F_{cent A}\frac{h}{2}sin(\alpha^{'})-R_{B}\frac{h}{2}sin(\alpha^{'})-F_{cent B}\frac{h}{2}cos(\alpha^{'})
=
0.
\label{equacao_equilibrio_momentos}
\end{equation}
The shear force depicted in Fig.\ref{fig:Esquematico_sistema_mecanico} by an $\vec{S}$, is decomposed in the $x$ and $y$ axes directions resulting in $S_{x}$ and $S_{y}$ presented in (\ref{equacao_S_x}) and (\ref{equacao_S_y}) respectively.
\begin{equation}
S_{x}
=
\vec{S}sin(\alpha^{'}).
\label{equacao_S_x}
\end{equation}
\begin{equation}
S_{y}
=
\vec{S}cos(\alpha^{'}).
\label{equacao_S_y}
\end{equation}
From this, the following required equation is the equilibrium of forces in the $y$ axis direction.
The obtained equation is disclosed in (\ref{equacao_equilibrio_forcas_y}). 
\begin{equation}
\Sigma F_{y}
=
R_{B}+F_{cent A}+S_{y}
=
0
\Leftrightarrow
R_{B}=-F_{cent A}-S_{y}.
\label{equacao_equilibrio_forcas_y}
\end{equation} 
The final required system equation is the $x$ axis direction forces equilibrium equation.
The resulting equation is shown in (\ref{equacao_equilibrio_forcas_x}). 
\begin{equation}
\Sigma F_{x}
=
-R_{A}-F_{cent B}+S_{x}
=
0
\Leftrightarrow
R_{A}=-F_{cent B}+S_{x}.
\label{equacao_equilibrio_forcas_x}
\end{equation} 
Substitution of the obtained relations (\ref{equacao_equilibrio_forcas_y}) and (\ref{equacao_equilibrio_forcas_x}) in the moment equilibrium equation (\ref{equacao_equilibrio_momentos}) will result in (\ref{equacao_equilibrio_momentos_subs_1}). 
\begin{equation}
-F_{cent B}\frac{h}{2}cos(\alpha^{'})+S_{x}\frac{h}{2}cos(\alpha^{'})+F_{cent A}\frac{h}{2}sin(\alpha^{'})+F_{cent A}\frac{h}{2}sin(\alpha^{'})+S_{y}\frac{h}{2}sin(\alpha^{'})-F_{cent B}\frac{h}{2}cos(\alpha^{'})
=
0.
\label{equacao_equilibrio_momentos_subs_1}
\end{equation}
From the relations (\ref{equacao_S_x}) and (\ref{equacao_S_y}), equation (\ref{equacao_equilibrio_momentos_subs_1}) turns to (\ref{equacao_equilibrio_momentos_subs_2}).
\begin{equation}
F_{cent A}hsin(\alpha^{'})-F_{cent B}hcos(\alpha^{'})+\vec{S}hsin(\alpha^{'})cos(\alpha^{'})
=
0.
\label{equacao_equilibrio_momentos_subs_2}
\end{equation} 
From this, the following step is presented in,
\begin{equation}
\frac{F_{cent B}cos(\alpha^{'})-F_{cent A}sin(\alpha^{'})}{sin(\alpha^{'})cos(\alpha^{'})}
=
\vec{S}.
\label{equacao_equilibrio_momentos_subs_3}
\end{equation}
According to equations (\ref{equacao_F_centrifuga}), (\ref{equacao_r_local}), (\ref{equacao_alfa_linha}) and (\ref{equacao_beta_linha}), the centrifugal forces in points ``A'' and ``B'' are defined in,
\begin{equation}
F_{cent A}
=
\frac{1}{4}\frac{U^{2}\rho V}{r_{avg}}\frac{\pi-2\alpha^{'}}{2\alpha^{'}},
\label{equacao_F_centrifuga_A}
\end{equation}
\begin{equation}
F_{cent B}
=
\frac{1}{4}\frac{U^{2}\rho V}{r_{avg}}\frac{2\alpha^{'}}{\pi-2\alpha^{'}}.
\label{equacao_F_centrifuga_B}
\end{equation}
The value of the shear force depicted by an $\vec{S}$ in Fig.\ref{fig:Esquematico_sistema_mecanico} is equal to a quarter of the total shear force acting upon the pre-transitional vortex defined in (\ref{equacao_F_shear}).
Therefore, $\vec{S}$ is defined in,
\begin{equation}
\vec{S}
=
\frac{1}{4}(\mu+\mu_{T}) SA,
\label{equacao_F_shear_quarter}
\end{equation}
As such, from these relations the resulting equation from (\ref{equacao_equilibrio_momentos_subs_3}) is disclose in,
\begin{equation}
\frac{U^{2}\rho V}{r_{avg}}\left[\frac{ \left(\frac{2\alpha^{'}}{\pi-2\alpha^{'}}\right) \cos\left(\alpha^{'}\right)-\left( \frac{\pi-2\alpha^{'}}{2\alpha^{'}} \right) \sin\left(\alpha^{'}\right) }{ \sin\left(\alpha^{'}\right)\cos\left(\alpha^{'}\right) }\right]
=
(\mu+\mu_{T}) SA.
\label{equacao_equilibrio_momentos_subs_4}
\end{equation}
From the mean linear velocity of the pre-transitional vortex defined in (\ref{equacao_U_avg}) and from the definition of area, ``A``, and volume, ``V``, defined in (\ref{equacao_area}) and (\ref{equacao_volume}) respectively, the ratio between pre-transitional vortex acting shear and centrifugal forces is given in (\ref{equacao_racio_abreviatura}). 
\begin{equation}
U
=
\omega r_{avg},
\label{equacao_U_avg}
\end{equation}
\begin{equation}
ratio
=
\frac{F_{shear}}{F_{cent}}=\frac{(\nu+\nu_{T})2S}{\left(\omega r_{avg}\right)^{2}}.
\label{equacao_racio_abreviatura}
\end{equation} 
The solution of this system in terms of angle $\alpha^{'}$ is then,
\begin{equation}
\left[\frac{ \left(\frac{2\alpha^{'}}{\pi-2\alpha^{'}}\right) \cos\left(\alpha^{'}\right)-\left( \frac{\pi-2\alpha^{'}}{2\alpha^{'}} \right) \sin\left(\alpha^{'}\right) }{ \sin\left(\alpha^{'}\right)\cos\left(\alpha^{'}\right) }\right]
=
ratio.
\label{equacao_racio}
\end{equation}
The evolution of ratio to $\alpha^{'}$ of the exact solution (\ref{equacao_racio}) is presented in Fig.\ref{fig:ratio_evolution}.
According to relation (\ref{equacao_racio}) and observing Fig.\ref{fig:ratio_evolution}, an unexpected evolution of the angle $\alpha^{'}$ is predicted by the mechanical model.
Based on (\ref{equacao_racio}), this solution of the mechanical model approximation used in this work predicted an increase of the angle $\alpha^{'}$ with increasing shear.
This in turn means that according to (\ref{equacao_uv_energia_explicita_resultado}) and (\ref{equacao_alfa_linha}), the calculated values of $\overline{u^{'}v^{'}}$ will be negative.
The $\alpha$ angle increases from $\frac{\pi}{2}$ to its limiting value $\pi$.
This also signifies that the correct pre-transitional vortex shape under the effect of shear is the one presented in Fig \ref{fig:Esquematico_deformacao_circulo_previsto}.
Since the relation (\ref{equacao_racio}) is far too complex to be use in a numerical transition closure, a simpler function is used instead.
The used function is a good approximation of the exact relation and is presented in,
\begin{equation}
tan\left[1.8\left(\alpha^{'}-\frac{\pi}{4}\right)^{0.7}\right]
=
ratio.
\label{equacao_racio_simplificada}
\end{equation}   
The used mechanical approximation function in the transition V-model numerical implementation is here presented in (\ref{equacao_racio_simplificada_formulacao}).
This is done in order to simplify implementation attempts by the reader. 

\begin{equation}
\alpha^{'}
=
min\left[ tan^{-1}\left(\frac{ratio}{1.8}\right)^{1.42857}+\frac{\pi}{4},\frac{\pi}{2}\right]. 
\label{equacao_racio_simplificada_formulacao}
\end{equation}

In (\ref{equacao_racio_simplificada_formulacao}), the minimum function is applied since the maximum value of $\alpha^{'}$ is $\frac{\pi}{2}$.
This correspond to a perfectly flat pre-transitional vortex deformation angle.
For values of $ratio=0$ the minimum value of $\alpha^{'}$ is $\frac{\pi}{4}$.
This represents an undeformed circular pre-transitional vortex.

\section{The transition V-model transport equation}
\label{sec:9}
The Reynolds stress transport equations were first closed by \cite{Rotta1951}, setting the foundations for second order turbulence models.
The transport equations are here presented in (\ref{equacao_transporte_tensoes_Reynolds}) as presented by \cite{Wilcox1994}.

\begin{equation}
\frac{\partial\tau_{ij}}{\partial t} + U_{k}\frac{\partial\tau_{ij}}{\partial x_{k}}
=
-\tau_{ik}\frac{\partial U_{j}}{\partial x_{k}}-\tau_{jk}\frac{\partial U_{i}}{\partial x_{k}} 
+ \epsilon_{ij}
- \Pi_{ij}
+\frac{\partial}{\partial x_{k}}\left[\nu\frac{\partial\tau_{ij}}{\partial x_{k}} + C_{ijk} \right], 
\label{equacao_transporte_tensoes_Reynolds}
\end{equation}

\begin{equation}
\epsilon_{ij}
=
\overline{2\mu\frac{\partial u^{'}_{i}}{\partial x_{k}} \frac{\partial u^{'}_{j}}{\partial x_{k}}}, 
\label{equacao_transporte_tensoes_Reynolds_epsilon}
\end{equation}

\begin{equation}
\Pi_{ij}
=
\overline{ p^{'}\left(\frac{\partial u^{'}_{i}}{\partial x_{j}}+ \frac{\partial u^{'}_{j}}{\partial x_{i}} \right)}, 
\label{equacao_transporte_tensoes_Reynolds_PIij}
\end{equation}

\begin{equation}
C_{ijk}
=
\overline{\rho u^{'}_{i}u^{'}_{j}u^{'}_{k}} + \overline{p^{'}u_{i}^{'}}\delta_{jk} + \overline{p^{'}u_{j}^{'}}\delta_{ik}.
\label{equacao_transporte_tensoes_Reynolds_Cijk}
\end{equation}

In (\ref{equacao_transporte_tensoes_Reynolds}), $\tau_{ij}=-\overline{\rho u_{i}^{'}u_{j}^{'}}$, represents the Reynolds stress tensor.
The presented terms on the right hand side of (\ref{equacao_transporte_tensoes_Reynolds}), are from left to right, two terms of production, one for turbulent dissipation presented in (\ref{equacao_transporte_tensoes_Reynolds_epsilon}), another for pressure-strain shown in (\ref{equacao_transporte_tensoes_Reynolds_PIij}) and finally the molecular and turbulent diffusion.
This last term is presented in (\ref{equacao_transporte_tensoes_Reynolds_Cijk}).
Considering the normal Reynolds stresses transport equations, that is, the trace components of the stress tensor, the number of terms is reduced resulting in (\ref{equacao_transporte_tensoes_Reynolds_ii}).
Through the condition imposed by the equation of continuity applied to turbulent velocity fluctuations the pressure-strain term is null.

\begin{equation}
\frac{\partial\tau_{ii}}{\partial t} + U_{k}\frac{\partial\tau_{ii}}{\partial x_{k}}
=
-2\tau_{ik}\frac{\partial U_{i}}{\partial x_{k}} 
+ \overline{2\mu\frac{\partial u^{'}_{i}}{\partial x_{k}} \frac{\partial u^{'}_{i}}{\partial x_{k}}}  
+\frac{\partial}{\partial x_{k}}\left[\nu\frac{\partial\tau_{ii}}{\partial x_{k}} + \overline{\rho u^{'}_{i}u^{'}_{i}u^{'}_{k}} + 2\overline{p^{'}u_{i}^{'}}\delta_{ik}  \right]. 
\label{equacao_transporte_tensoes_Reynolds_ii}
\end{equation}

Since $k_{i}=\frac{1}{2}\overline{u_{i}^{'}u_{i}^{'}}$, considering then the relation that, $\tau_{ii}=-2\rho k_{i}$, we obtain the following transport equation for turbulent kinetic energy in (\ref{equacao_transporte_tensoes_Reynolds_ki}).

\begin{equation}
\frac{\partial k_{i}}{\partial t} + U_{k}\frac{\partial k_{i}}{\partial x_{k}}
=
-\overline{u_{i}^{'}u_{k}^{'}}\frac{\partial U_{i}}{\partial x_{k}} 
- \overline{\nu\frac{\partial u^{'}_{i}}{\partial x_{k}} \frac{\partial u^{'}_{i}}{\partial x_{k}}}  
+\frac{\partial}{\partial x_{k}}\left[\nu\frac{\partial k_{i}}{\partial x_{k}} - \frac{1}{2}\overline{\rho u^{'}_{i}u^{'}_{i}u^{'}_{k}} - \overline{p^{'}u_{i}^{'}}\delta_{ik}  \right]. 
\label{equacao_transporte_tensoes_Reynolds_ki}
\end{equation}

The proposed transition model pre-transition turbulent kinetic energy transport equation components are based on some terms of equation (\ref{equacao_transporte_tensoes_Reynolds_ki}).
The transition closure production term was obtained by analyzing the first term in (\ref{equacao_transporte_tensoes_Reynolds_ki}).
This analysis will be performed considering a flat-plate flow far from its leading edge region.
Within the shear region of the flow, $\overline{u_{1}^{'}u_{2}^{'}}$ or, $\overline{u^{'}v^{'}}$, will be negative.
The shear value of $\partial U_{1}/\partial x_{2}$ will be positive.
Thus this term is a production term in mean flow shear conditions.
The mean flow shear magnitude, $S$, is used as the mean flow property that drives pre-transitional turbulence production.
It must be noted that for clarity reasons, a term was omitted in the previously presented $\overline{u^{'}v^{'}}$ calculation in (\ref{equacao_uv_energia_explicita_resultado}).
The term is a function responsible for the evaluation of the mean flow and turbulent scales proximity, $F_{S}$.
This is defined in (\ref{equation_F_S}).
\begin{equation}
F_{S}
=
1-min\left(mag\left(\frac{S_{size}-r_{avg}}{max\left(S_{size},r_{avg}\right)}\right),1\right)
\label{equation_F_S}
\end{equation} 
The used  mean flow shear scale function, $S_{size}$, is defined in (\ref{equation_S_size}).
\begin{equation}
S_{size}
=
\sqrt{\frac{\nu}{S}}
\label{equation_S_size}
\end{equation}
The $\overline{u^{'}v^{'}}$ computation is then performed according to,  
\begin{equation}
\overline{u^{'}v^{'}}
=
\frac{ \left(r_{avg}\omega\right)^{2} \cos\left(\alpha\right)}{2} F_{S}.
\label{equacao_uv_energia_explicita_resultado_final}
\end{equation}
The production term used is presented in,
\begin{equation}
Prod_{k_{p}}
=
-C_{P_{k}} S \overline{u^{'}v^{'}}.
\label{Termo_producao}
\end{equation}
In the latter equation, $C_{P_{k}}$ is the V-model calibration constant equal to $0.835$.

The second term on the RHS of (\ref{equacao_transporte_tensoes_Reynolds_ki}) will be used to obtain the dissipation term of the presented transition model transport equation.
Calculation of the cross partial derivatives from the latter term reveals a fundamental result.
In order to simplify the partial derivatives calculation, the following assumption is required.
The equations (\ref{equacao_x}) and (\ref{equacao_y}) definitions impose a phase shift between $x^{'}$ and $y^{'}$.
For an undeformed pre-transitional vortex, implying $\alpha=\frac{\pi}{2}$, and summing a phase shift value of $\frac{\pi}{4}$ to both equations, (\ref{equacao_x}) and (\ref{equacao_y}), we obtain the same phase shift imposition through (\ref{equacao_x_alterado}) and (\ref{equacao_y_alterado}) respectively.
A detailed explanation of the process is presented in (\ref{equacao_x_alterado_explicacao}) and (\ref{equacao_y_alterado_explicacao}).
\begin{equation}
x^{'}
=
-r_{avg} \cos\left(\omega t+\frac{\frac{\pi}{2}}{2}+\frac{\pi}{4} \right)
=
r_{avg} \sin\left(\omega t\right),
\label{equacao_x_alterado_explicacao}
\end{equation}
\begin{equation}
y^{'}
=
-r_{avg} \cos\left(\omega t-\frac{\frac{\pi}{2}}{2}+\frac{\pi}{4} \right)
=
-r_{avg} \cos\left(\omega t\right).
\label{equacao_y_alterado_explicacao}
\end{equation}
\begin{equation}
x^{'}
=
r_{avg} \sin\left(\omega t\right),
\label{equacao_x_alterado}
\end{equation}

\begin{equation}
y^{'}
=
-r_{avg} \cos\left(\omega t\right).
\label{equacao_y_alterado}
\end{equation}
The new velocity fluctuations $u^{'}$ and $v^{'}$ are then represented by (\ref{equacao_u_alterado}) and (\ref{equacao_v_alterado}) respectively.

\begin{equation}
u^{'}
=
r_{avg} \omega \cos\left(\omega t\right),
\label{equacao_u_alterado}
\end{equation}

\begin{equation}
v^{'}
=
r_{avg} \omega \sin\left(\omega t\right).
\label{equacao_v_alterado}
\end{equation}
Considering then a cross partial derivative where the length scale of $\partial y$ is comparable to the length scale of $\partial y^{'}$ resulting in, 

\begin{equation}
\frac{\partial u^{'}}{\partial y}
\approx
\frac{\partial u^{'}}{\partial y^{'}}
=
\frac{\partial r_{avg} \omega \cos\left(\omega t\right)}{\partial y^{'}}.
\label{derivada_cruzada}
\end{equation} 
It can be observed that apparently the latter partial derivative can not be easily solved.
However, the $\omega t$ dependence can be written as,

\begin{equation}
y^{'}
=
-r_{avg} \cos\left(\omega t\right)
\Leftrightarrow
\omega t
=
arcos\left(\frac{-y^{'}}{r_{avg}}\right).
\label{wt}
\end{equation}
By doing so the cross partial derivative becomes then,

\begin{equation}
\frac{\partial u^{'}}{\partial y^{'}}
=
\frac{\partial r_{avg} \omega \cos\left( arcos\left(\frac{-y^{'}}{r_{avg}}\right) \right)}{\partial y^{'}}
=
\frac{\partial \omega r_{avg} \frac{-y^{'}}{r_{avg}}}{\partial y^{'}}.
\label{derivada_cruzada_sem_wt}
\end{equation}
The resulting cross partial derivative is presented in (\ref{derivada_cruzada_sem_wt_final}).
\begin{equation}
\frac{\partial u^{'}}{\partial y^{'}}
=
-\frac{\partial \omega y^{'} }{\partial y^{'}}
=
-\omega.
\label{derivada_cruzada_sem_wt_final}
\end{equation}
Therefore the cross partial derivative of the dissipation term (\ref{equacao_transporte_tensoes_Reynolds_epsilon}) is equal to a form of rotational velocity of the undeformed pre-transitional turbulent vortices, $\omega$.
It must be noted that due to the assumption of $\partial y \approx \partial y^{'}$, a scale proximity between turbulent and mean flow must exist.
The non-cross partial derivatives of (\ref{equacao_transporte_tensoes_Reynolds_epsilon}) were not considered.
The resulting dissipation term in the V-model transport equation is then presented in (\ref{VSA-transport-equation-termo_dissipacao}).
It must be noted that although the cross partial derivative, (\ref{derivada_cruzada_sem_wt_final}), of the dissipation term, (\ref{equacao_transporte_tensoes_Reynolds_epsilon}), has a negative sign, the dissipation term makes use of the square of this cross partial derivative.
\begin{equation}
Dest_{k_{p}}
=
\nu\Omega^{2}F_{\Omega}
\approx
\overline{\nu\frac{\partial u^{'}_{i}}{\partial x_{k}} \frac{\partial u^{'}_{i}}{\partial x_{k}}}.
\label{VSA-transport-equation-termo_dissipacao}
\end{equation} 
Similar to the used ''shear-sheltering'' effect function presented in the work of \cite{Walters2008}, mean flow vorticity, $\Omega$, is used as a form of rotational velocity instead of the pre-transitional vortex rotational speed, $\omega$, for the destruction effect within the boundary layer.
Citing from the work of \cite{Walters2008}, ``Shear-sheltering refers to the damping of turbulence dynamics that occurs in thin regions of high vorticity...''.
This is also done in order to allow the mean flow characteristics to control transition onset as depicted in Fig.\ref{fig:Esquematico_V_SA}.
Also, mean flow vorticity can be related to stabilization effects of turbulence, which can be interpreted as a turbulence sink.
Such can be observed in relaminarization experiments of turbulent flow inside tube coils as presented by the work of \cite{Viswanath1978}.
This was also calculated with DNS, \cite{Noorani2013}.
The $F_{\Omega}$ function assures the aforementioned assumption of $\partial y \approx \partial y^{'}$.
$F_{\Omega}$ can also be interpreted as a measure of the mean and turbulent flow scale proximity.
This is defined in (\ref{equation_F_Omega}).
\begin{equation}
F_{\Omega}
=
1-min\left(mag\left(\frac{\Omega_{size}-r_{avg}}{max\left(\Omega_{size},r_{avg}\right)}\right),1\right).
\label{equation_F_Omega}
\end{equation}
The used  mean flow vorticity scale function, $\Omega_{size}$, is defined in (\ref{equation_Omega_size}).
\begin{equation}
\Omega_{size}
=
\sqrt{\frac{\nu}{\Omega}}
\label{equation_Omega_size}
\end{equation}
The turbulent diffusion component of the transport equation was chosen to be equal to a common turbulent kinetic energy transport equation such as the model of \cite{Craft1996}.
The resulting pre-transitional turbulent kinetic energy transport equation is then presented in,
\begin{equation}
\frac{Dk_{p}}{Dt}
=
Prod_{k_{p}}
-Dest_{k_{p}}
+\frac{\partial}{\partial x_{j}}\left[\left(\nu+\nu_{T}\right)\frac{\partial k_{p}}{\partial x_{j}}\right].
\label{VSA-transport-equation-k}
\end{equation}
From the transition V-model, the pre-transition turbulent kinetic energy will apply an induced viscosity defined in (\ref{VSA-turb_viscosity}).
\begin{equation}
\nu_{Tuv}
=
\frac{-\overline{u^{'}v^{'}}}{mag \left( \nabla U \right)}.
\label{VSA-turb_viscosity}
\end{equation}
This will then be the small effect of the pre-transitional turbulent kinetic energy on the viscosity within the pre-transitional region of the laminar boundary layer.

\section{Transition V-model and Spalart-Allmaras turbulence model coupling}

The Spalart-Allmaras is a well known one-equation turbulence model.
It was first presented by \cite{Spalart1994}.
As stated before, the transition V-model was coupled to the Spalart-Allmaras turbulence model.
This was performed by simply adding a control function in the production term of the turbulence model.
This control function depends upon the ratio value obtained by the mechanical model approximation in (\ref{equacao_racio_abreviatura}).
The performed implementation is shown in (\ref{VSA-transport-equation-nuTilda}).

\begin{equation}
\frac{D\tilde{\nu_{T}}}{Dt}
=
C_{b1}\tilde{S}\tilde{\nu_{T}}exp\left(-C_{\tilde{\nu_{T}}}ratio\right)
-C_{w1}f_{w}\left(\frac{\tilde{\nu_{T}}}{d}\right)^{2}
+\frac{\partial}{\partial x_{j}}\left[\left(\tilde{\nu_{T}}+\frac{\nu}{\sigma}\right)\frac{\partial \tilde{\nu_{T}}}{\partial x_{j}}\right].
\label{VSA-transport-equation-nuTilda}
\end{equation}

As can be seen in the first component of the RHS of (\ref{VSA-transport-equation-nuTilda}), the control function is an exponential term.
The $C_{\tilde{\nu_{T}}}$ constant is equal to $0.5$.
Also in order to account for the effect of the pre-transitional viscosity, (\ref{VSA-turb_viscosity}), the Spalart-Allmaras turbulence model production term is slightly changed to,
\begin{equation}
\tilde{S}
=
\frac{\Omega}{\sqrt{2}}+\frac{\tilde{\nu_{T}}}{\left(\kappa d\right)^{2}}f_{v2}.
\label{equation_SA_production}
\end{equation}
For brevity's sake, further description of the Spalart-Allmaras model is omitted here.
The remaining Spalart-Allmaras model is briefly described in \cite{Rumsey2001}.

The total turbulent viscosity that the V-SA transition model predicts is a sum of the obtained turbulent viscosity from the SA model and the transition V-model.
This is presented in (\ref{VSA-total_turb_viscosity}).
\begin{equation}
\nu_{T}
=
\nu_{Tuv}+
f_{v1}\tilde{\nu_{T}}.
\label{VSA-total_turb_viscosity}
\end{equation}

\section{Results and Discussion}
\label{sec:10}
All of the obtained results for the V-SA and SA models were calculated using the open-source software OpenFoam.
The considered cases were all steady-state and incompressible.
These were performed with a constant density of $\rho=1.2\left(kg/m^{3}\right)$.
The results calculated with OpenFOAM were run with a pressure based solver SIMPLE, linear discretization for laplacian terms 
and LUST discretization scheme for all possible divergence terms. In OpenFOAM the divergence term, (\ref{Termo_OpenFOAM}), requires a linear discretization. 

\begin{equation}
div((nuEff*dev(grad(U).T()))).
\label{Termo_OpenFOAM}
\end{equation}  

All of the obtained results with the empirical transition model $\gamma-R_{e\theta}$ were computed using Ansys Fluent 13.0.
The used discretization scheme was the second order upwind, SOU, and for pressure the linear setting was applied. 

The V-SA wall boundary conditions for all tested cases were Dirichlet conditions for pre-transition turbulent kinetic energy, that is, $k_{p}=0$ at the wall. 

As stated in the Introduction section, the first comparison is performed by using the results from ERCOFTAC obtained by the experimental work of \cite{J.1990a}, for one of the zero-pressure-gradient, ZPG, flat-plate test cases.
The tested case is then the T3A.
The upstream conditions for this test case is presented in table \ref{tab:ERCOFTAC_upstream}.

\begin{table}
\caption{ERCOFTAC ZPG Flat-Plate Upstream Condition}
\label{tab:ERCOFTAC_upstream}
\begin{centering}
{\footnotesize }\begin{tabular}{|c|c|c|c|}
\hline 
{\footnotesize $Case$} & {\footnotesize $Tu (\%)$} & {\footnotesize $U (m/s)$} \tabularnewline
\hline 
{\footnotesize $T3A$} & {\footnotesize $3.0$} & {\footnotesize $5.4$} \tabularnewline
\hline
\end{tabular}
\par\end{centering}
\end{table}

For the zero pressure gradient test case the flat-plate mesh used was structured and had $y^+$ values below $0.1$.
The flat-plate had $1.7$ meters of extension with $200$ mesh points over its surface.
These clustered near the leading edge of the plate.
The leading edge had a curvature radius of $0.002$ meters.
Along the leading edge the mesh had $30$ mesh points.
The wall perpendicular spacing of the first layer of cells over the flat-plate was $1\times10^{-5}$ meters.
The velocity inlet was located at $0.15$ meters from the leading edge.
This short extension had $110$ mesh nodes.
The top surface was located at $0.15$ meters above the flat-plate.
This vertical length had $110$ mesh points along it.

The inlet boundary conditions for this flat-plate ZPG ERCOFTAC test case is presented in table \ref{tab:ERCOFTAC ZPG Inlet Boundary Condition}.

\begin{table}
\caption{ERCOFTAC ZPG Flat-Plate Inlet Boundary Conditions}
\label{tab:ERCOFTAC ZPG Inlet Boundary Condition}
\begin{centering}
{\footnotesize }\begin{tabular}{|c|c|c|c|}
\hline 
{\footnotesize $Case$} & {\footnotesize $k_{p} (m^{2}/s^{2})$} & {\footnotesize $U (m/s)$} & {\footnotesize $\tilde{\nu_{T}}/\nu$} \tabularnewline
\hline 
{\footnotesize $T3A$} & {\footnotesize $0.039366$} & {\footnotesize $5.4$} & {\footnotesize $3$} \tabularnewline
\hline
\end{tabular}
\par\end{centering}
\end{table}

\begin{figure}
	\centering
			\includegraphics[width=0.9\textwidth]{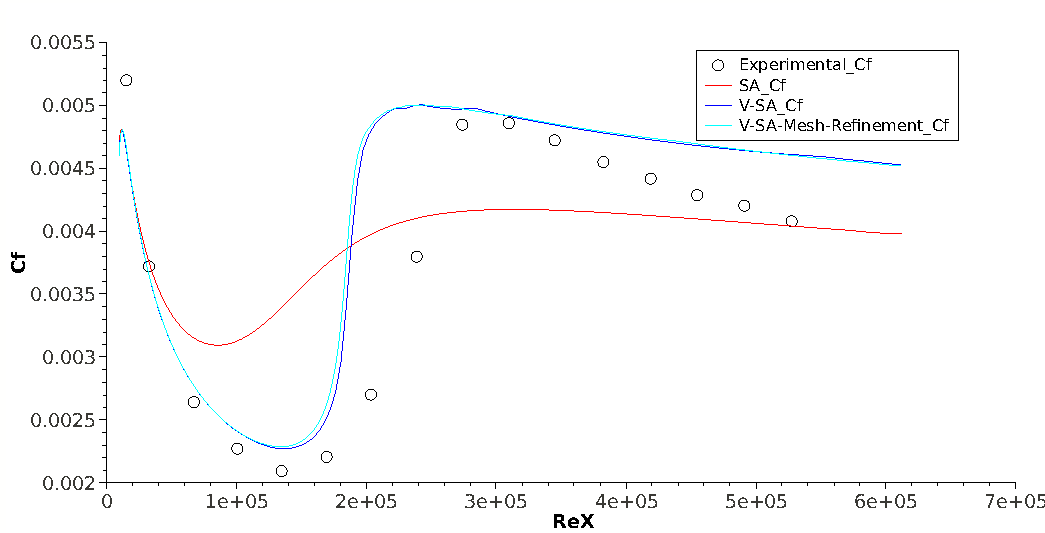}
	
	\caption{Comparison of experimental data from ERCOFTAC T3A flat-plate test case skin-friction coefficient distribution with the non-transition SA closure and the transition V-SA model.
The used structured computational grids for the mesh refinement study had the respective sizes of $110\times340$ and $220\times680$ nodes.}
\label{fig:CF_T3A}
\end{figure}

The results for the flat-plate T3A ERCOFTAC test case are presented in Fig.\ref{fig:CF_T3A}.
These include a mesh refinement validation.
This was performed using an equal mesh with double number of nodes over the entire computational domain.

\begin{figure}
	\centering
			\includegraphics[width=0.9\textwidth]{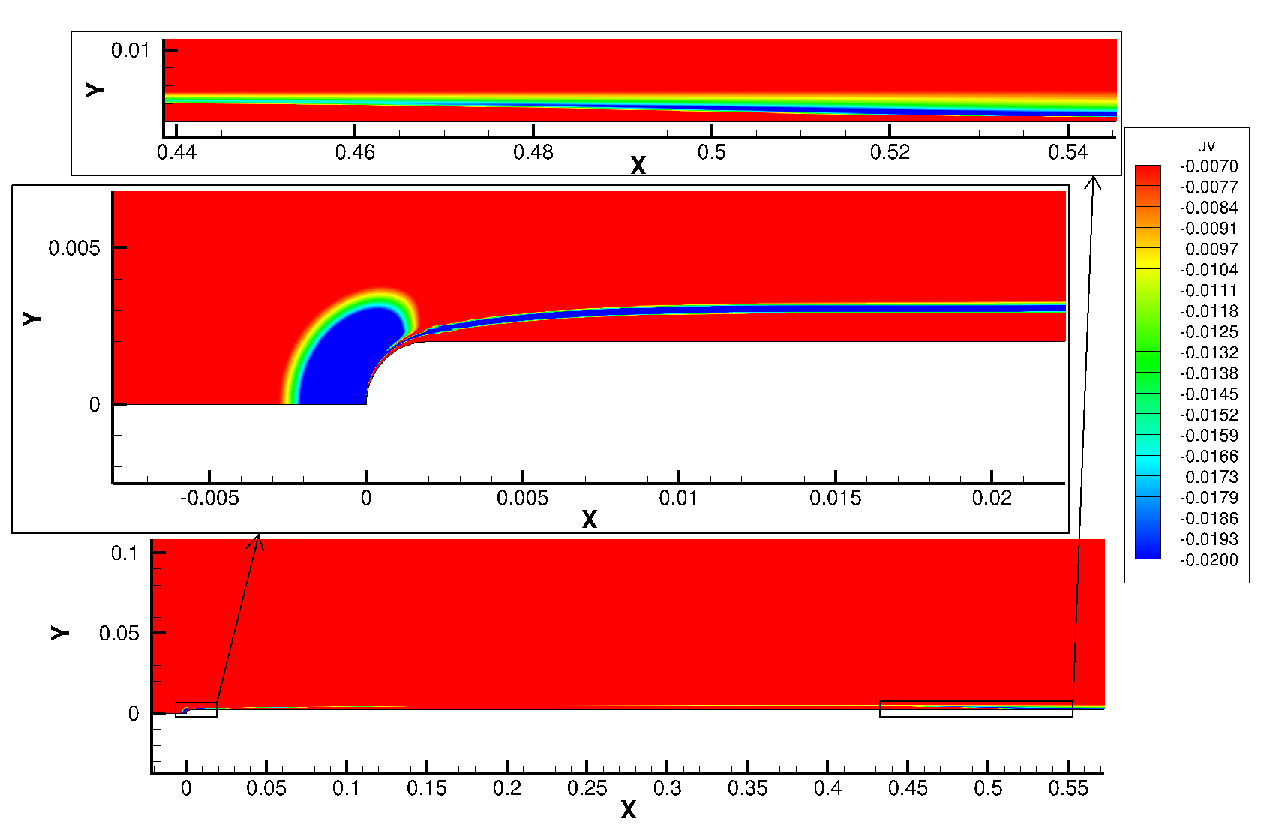}
	
	\caption{Contour map of $\overline{u^{'}v^{'}}$ values with detailed results near the leading edge and transition onset zone of the T3A flat-plate test case.}
\label{fig:uv_total_plane}
\end{figure}

As can be observed, although the Spalart-Allmaras model is able to calculate a certain transition, it predicts transition onset too early.
The V-SA transition model improves on the transition onset prediction, demonstrating a more correct behavior.
The V-SA mesh refinement study indicates that the results are mesh independent.
As already mentioned, the V-model transition closure calculates small pre-transitional negative values of $\overline{u^{'}v^{'}}$.
This is shown for the region near the leading edge and for the transition onset zone of the T3A test case in Fig.\ref{fig:uv_total_plane}.
As can be seen, when the mean flow characteristics are ideal the V-model predicts the piercing of the laminar boundary layer by the pre-transitional $\overline{u^{'}v^{'}}$.
This then activates the SA turbulence model production term.  

\begin{figure}
	\centering
			\includegraphics[width=0.9\textwidth]{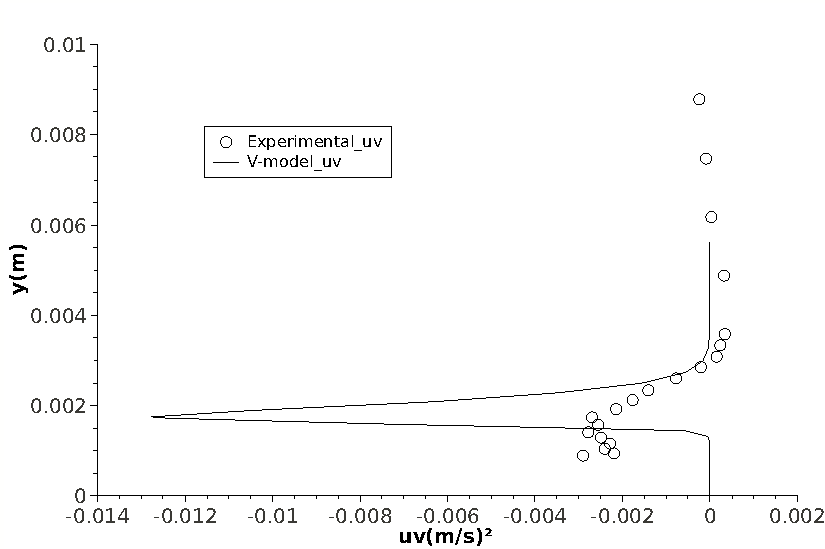}
	
	\caption{Comparison of ERCOFTAC flat-plate T3A experimental $\overline{u^{'}v^{'}}$ values with those predicted by the transition V-model in the axial position of $0.095$ meters or Rex of $3.24\times10^{4}$.
The negative pre-transitional $\overline{u^{'}v^{'}}$ values were related to ''splat-mechanism`` or ''inactive-motion`` by \cite{Bradshaw1994}.}
\label{fig:uv_0_095}
\end{figure}

\begin{figure}
	\centering
			\includegraphics[width=0.9\textwidth]{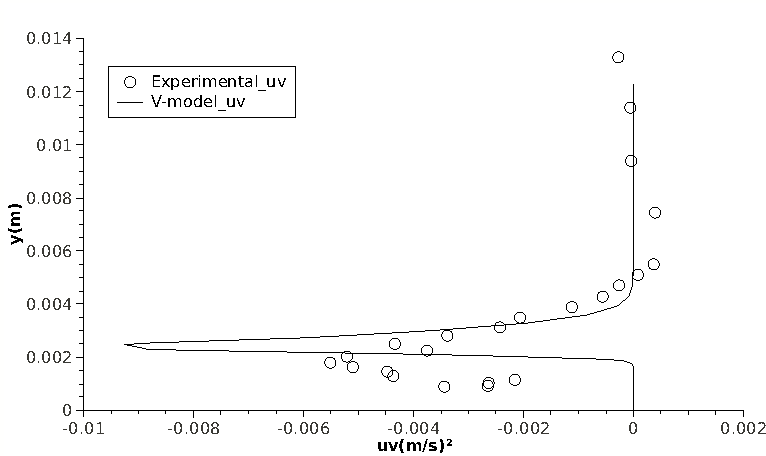}
	
	\caption{Comparison of ERCOFTAC flat-plate T3A experimental $\overline{u^{'}v^{'}}$ values with those predicted by the transition V-model in the axial position of $0.195$ meters or Rex of $6.70\times10^{4}$.}
\label{fig:uv_0_195}
\end{figure}

\begin{figure}
	\centering
			\includegraphics[width=0.9\textwidth]{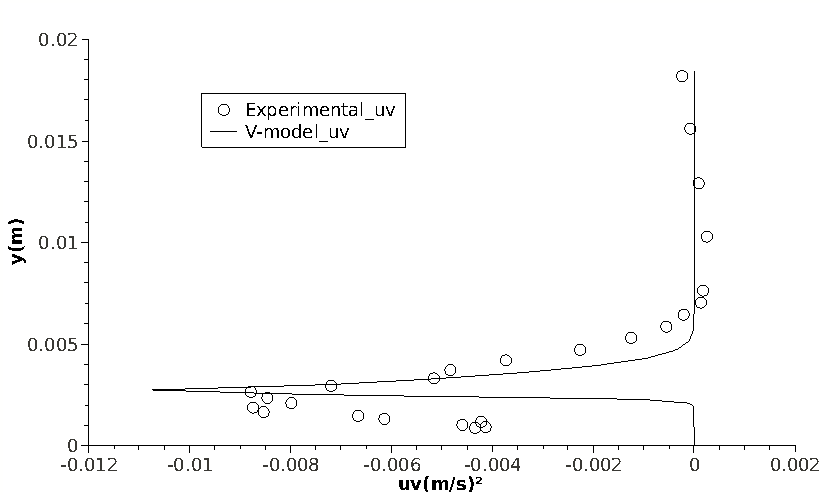}
	
	\caption{Comparison of ERCOFTAC flat-plate T3A experimental $\overline{u^{'}v^{'}}$ values with those predicted by the transition V-model in the axial position of $0.295$ meters or Rex of $10.06\times10^{4}$.}
\label{fig:uv_0_295}
\end{figure}

\begin{figure}
	\centering
			\includegraphics[width=0.9\textwidth]{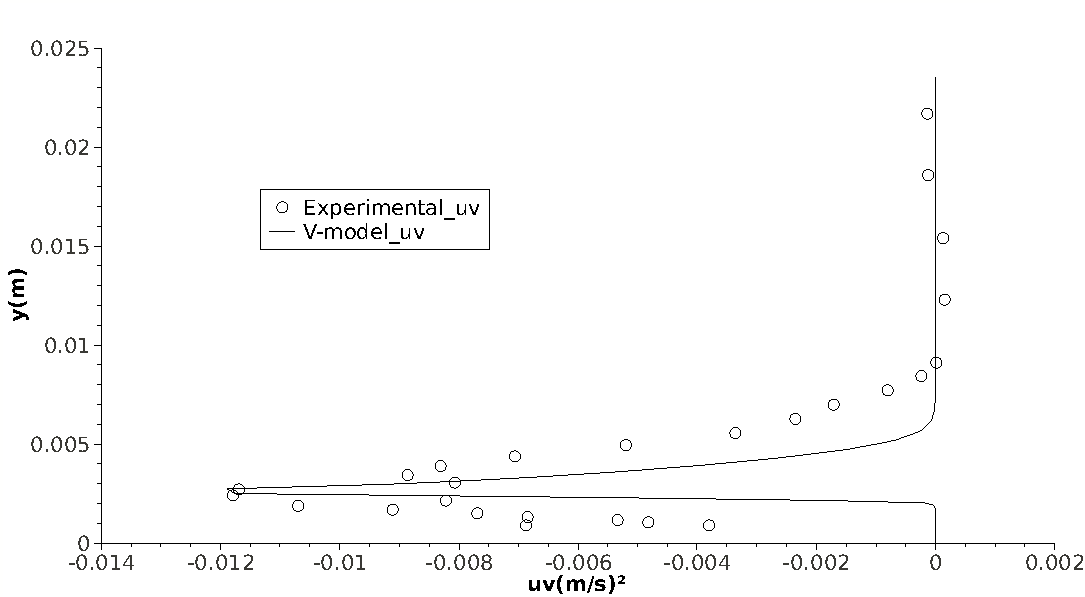}
	
	\caption{Comparison of ERCOFTAC flat-plate T3A experimental $\overline{u^{'}v^{'}}$ values with those predicted by the transition V-model in the axial position of $0.395$ meters or Rex of $13.48\times10^{4}$.}
\label{fig:uv_0_395}
\end{figure}

\begin{figure}
	\centering
			\includegraphics[width=0.9\textwidth]{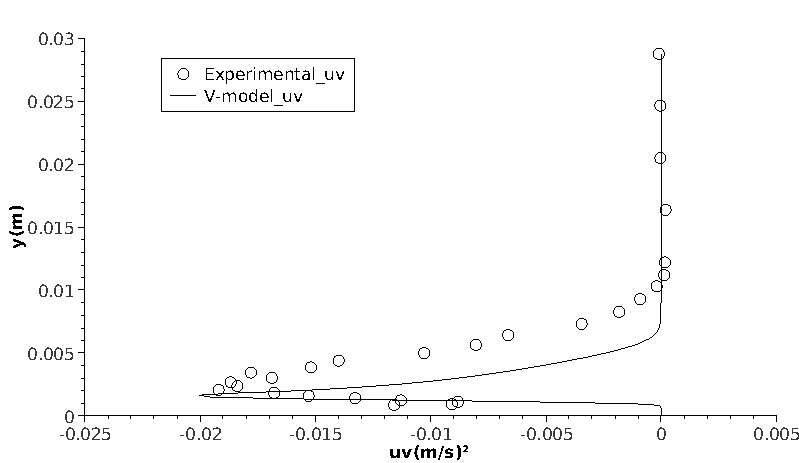}
	
	\caption{Comparison of ERCOFTAC flat-plate T3A experimental $\overline{u^{'}v^{'}}$ values with those predicted by the transition V-model in the axial position of $0.495$ meters or Rex of $16.92\times10^{4}$.}
\label{fig:uv_0_495}
\end{figure}

\begin{figure}
	\centering
			\includegraphics[width=0.9\textwidth]{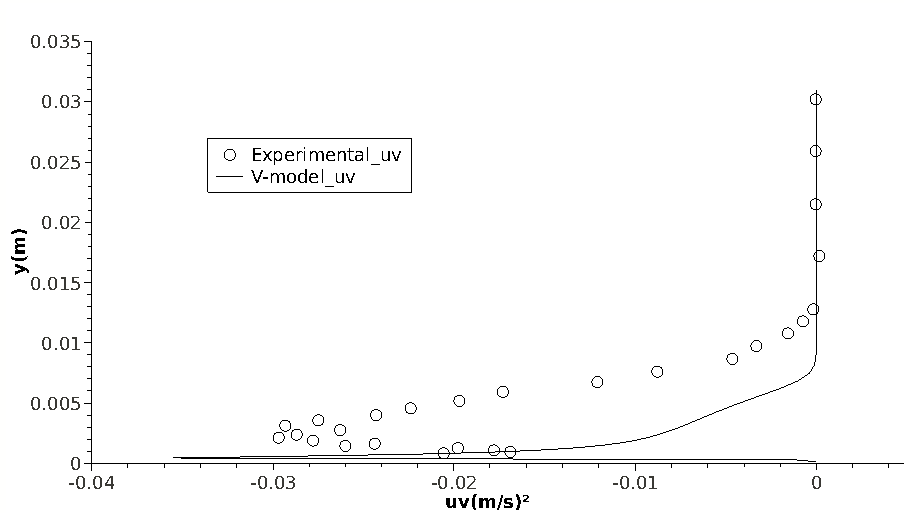}
	
	\caption{Comparison of ERCOFTAC flat-plate T3A experimental $\overline{u^{'}v^{'}}$ values with those predicted by the transition V-model in the axial position of $0.595$ meters or Rex of $20.35\times10^{4}$.}
\label{fig:uv_0_595}
\end{figure}

The proposed mechanical model approximation for pre-transitional vortex deformation predicts the distribution of $\overline{u^{'}v^{'}}$ in the pre-transitional boundary layer region.
A comparison is performed between the experimental ERCOFTAC database of $\overline{u^{'}v^{'}}$ and the V-model predicted values.
These comparisons are performed from the leading edge to the transition onset point.
The last comparison is one station in the middle of the transition region.
The results are presented in Figs.\ref{fig:uv_0_095}, \ref{fig:uv_0_195}, \ref{fig:uv_0_295}, \ref{fig:uv_0_395}, \ref{fig:uv_0_495} and \ref{fig:uv_0_595}.
These represent the axial positions in meters corresponding to $0.095$, $0.195$, $0.295$, $0.395$, $0.495$ and the transition section $0.595$ respectively.
The latter axial positions correspond to the Reynolds numbers of $3.24\times10^{4}$, $6.70\times10^{4}$, $10.06\times10^{4}$, $13.48\times10^{4}$, $16.92\times10^{4}$ and the transition section $20.35\times10^{4}$ respectively.

As can be seen, the V-model predicts an overshoot of $\overline{u^{'}v^{'}}$ near the leading edge.
However, the results greatly improve along the pre-transitional region.
The results in Fig.\ref{fig:uv_0_595}, corresponding to a station in the middle of the transition length region, show that the model predictions deviate from the measured values.

\begin{figure}
	\centering
			\includegraphics[width=0.9\textwidth]{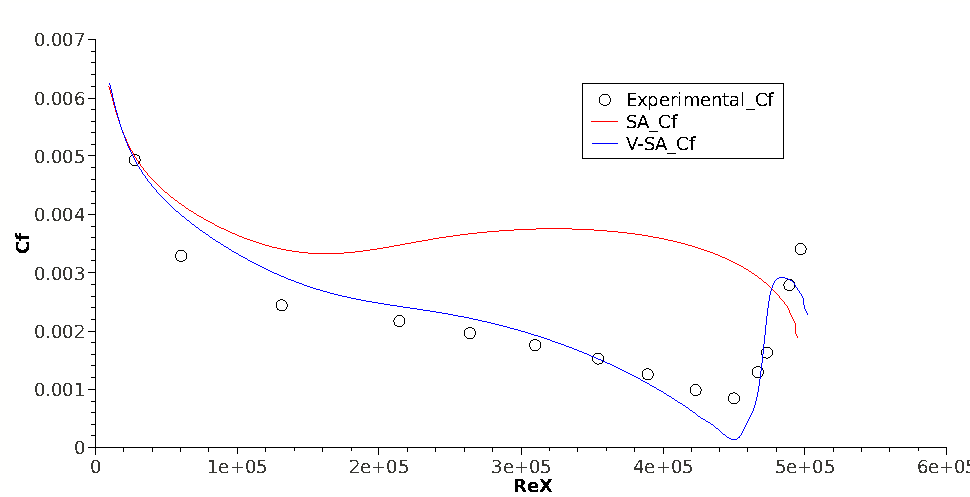}
	
	\caption{Comparison of experimental data from ERCOFTAC T3C3 flat-plate test case skin-friction coefficient distribution with the non-transition SA closure and the transition V-SA model.}
\label{fig:CF_T3C3}
\end{figure}

For the T3C test case the ERCOFTAC flat-plate pressure gradient transition experimental data from \cite{J.1990} was used.
The tested case was the T3C3.
The upstream conditions for this test case are presented in table \ref{tab:ERCOFTAC_T3C_upstream}.
\begin{table}
\caption{ERCOFTAC Pressure Gradient Flat-Plate Upstream Conditions}
\label{tab:ERCOFTAC_T3C_upstream}
\begin{centering}
{\footnotesize }\begin{tabular}{|c|c|c|c|}
\hline 
{\footnotesize $Case$} & {\footnotesize $Tu (\%)$} & {\footnotesize $U (m/s)$} \tabularnewline
\hline 
{\footnotesize $T3C3$} & {\footnotesize $3.0$} & {\footnotesize $3.7$} \tabularnewline
\hline
\end{tabular}
\par\end{centering}
\end{table}
The experimental pressure gradient flat-plate test case was performed with a structured mesh.
The bottom part of it was equal to the ZPG T3A test case mesh.
The upper component of the mesh was a curved wall surface designed to obtained the experimentally measured free-stream velocity variations.
The curved surface had $200$ mesh points along it.
The mesh points cluster near this surface and the spacing of the first layer of cells was $1\times10^{-4}$ meters.
This top section of the mesh had $30$ nodes along its vertical connectors.
Summing this with the previous bottom part of the mesh makes a total of $140$ mesh points in the vertical direction over the flat-plate.
The inlet of the complete mesh had a vertical length of $0.3$ meters.
The inlet boundary conditions for the T3C3 ERCOFTAC test case are presented in table \ref{tab:ERCOFTAC pressure gradient Inlet Boundary Conditions}.
The results for the flat-plate T3C3 ERCOFTAC test case are presented in Fig.\ref{fig:CF_T3C3}.
\begin{table}
\caption{ERCOFTAC Pressure Gradient Flat-Plate Inlet Boundary Conditions}
\label{tab:ERCOFTAC pressure gradient Inlet Boundary Conditions}
\begin{centering}
{\footnotesize }\begin{tabular}{|c|c|c|c|}
\hline 
{\footnotesize $Case$} & {\footnotesize $k_{p} (m^{2}/s^{2})$} & {\footnotesize $U (m/s)$} & {\footnotesize $\tilde{\nu_{T}}/\nu$} \tabularnewline
\hline 
{\footnotesize $T3C3$} & {\footnotesize $0.01848$} & {\footnotesize $3.7$} & {\footnotesize $3$} \tabularnewline
\hline
\end{tabular}
\par\end{centering}
\end{table}
As can be observed, the V-SA model's transition onset prediction is in accordance with the experimental data.
The SA turbulence model predicts an early transition onset.

\begin{figure}
	\centering
			\includegraphics[width=0.9\textwidth]{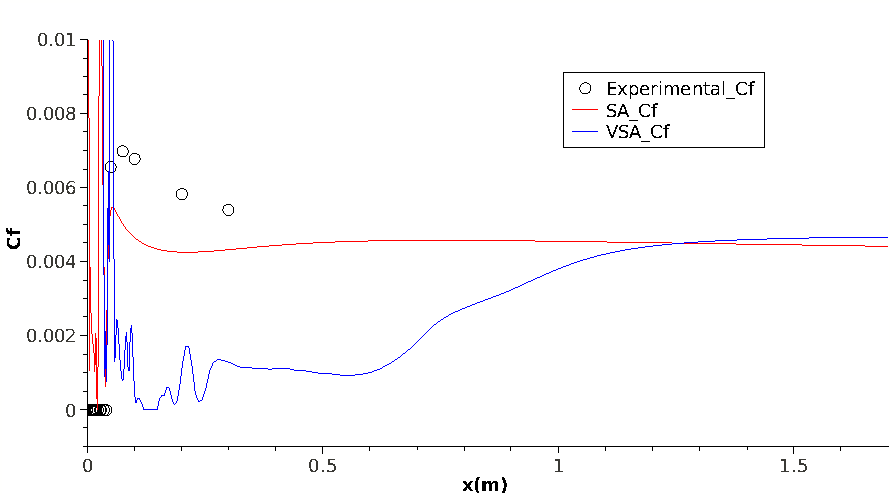}
	
	\caption{Comparison of experimental data from ERCOFTAC T3L1 flat-plate test case skin-friction coefficient distribution with the non-transition SA closure and the transition V-SA model.
The flat-plate leading edge oscillations of skin-friction coefficient are due to flow separation.}
\label{fig:CF_T3L1}
\end{figure}

As mentioned earlier, the ERCOFTAC T3L test cases of separation induced transition were considered for validation purposes.
The importance of such validation is made clear in the work of \cite{Hadzic1999}.
The experimental results from \cite{Coupland1996}, were used for the present validation. 
The used computational mesh was structured and had $y^+$ values below $0.1$ along the whole flat-plate surface.
The flat-plate had $1.7$ meters of extension with $200$ mesh points over its surface.
These clustered near the leading edge of the plate.
The leading edge had a curvature radius of $0.005$ meters, which is in accordance with the experimental setup.
Along the leading edge the mesh had $60$ mesh points.
The wall perpendicular spacing of the first layer of cells over the flat-plate was $1\times10^{-5}$ meters.
The velocity inlet was located at $0.15$ meters from the leading edge.
This short extension had $110$ mesh nodes.
The top surface was located at $0.15$ meters above the flat-plate.
This vertical length had $110$ mesh points along it.
The upstream conditions for the considered test cases are presented in table \ref{tab:ERCOFTAC_T3L_upstream}.
The inlet boundary conditions for the flat-plate ERCOFTAC T3L test cases are presented in table \ref{tab:ERCOFTAC T3L Inlet Boundary Conditions}.
\begin{table}
\caption{ERCOFTAC T3L Flat-Plate Upstream Conditions}
\label{tab:ERCOFTAC_T3L_upstream}
\begin{centering}
{\footnotesize }\begin{tabular}{|c|c|c|c|}
\hline 
{\footnotesize $Case$} & {\footnotesize $Tu (\%)$} & {\footnotesize $U (m/s)$} \tabularnewline
\hline 
{\footnotesize $T3L1$} & {\footnotesize $0.2$} & {\footnotesize $5.0$} \tabularnewline
\hline
{\footnotesize $T3L3$} & {\footnotesize $2.3$} & {\footnotesize $5.0$} \tabularnewline
\hline
{\footnotesize $T3L5$} & {\footnotesize $2.3$} & {\footnotesize $2.5$} \tabularnewline
\hline
\end{tabular}
\par\end{centering}
\end{table}
\begin{table}
\caption{ERCOFTAC T3L Flat-Plate Inlet Boundary Conditions}
\label{tab:ERCOFTAC T3L Inlet Boundary Conditions}
\begin{centering}
{\footnotesize }\begin{tabular}{|c|c|c|c|}
\hline 
{\footnotesize $Case$} & {\footnotesize $k_{p} (m^{2}/s^{2})$} & {\footnotesize $U (m/s)$} & {\footnotesize $\tilde{\nu_{T}}/\nu$} \tabularnewline
\hline 
{\footnotesize $T3L1$} & {\footnotesize $0.00015$} & {\footnotesize $5.0$} & {\footnotesize $5$} \tabularnewline
\hline
{\footnotesize $T3L3$} & {\footnotesize $0.019838$} & {\footnotesize $5.0$} & {\footnotesize $5$} \tabularnewline
\hline
{\footnotesize $T3L5$} & {\footnotesize $0.004959$} & {\footnotesize $2.5$} & {\footnotesize $5$} \tabularnewline
\hline
\end{tabular}
\par\end{centering}
\end{table}
For these cases, separation induced transition will be analyzed.
As such, the fluid kinematic viscosity used was the experimental value of $\nu=1.6\times10^{-5}\left(m^{2}/s\right)$.
As shown in table \ref{tab:ERCOFTAC_T3L_upstream}, the T3L1 test case represents the lower bound of turbulence intensity.
The results for this case are presented in Fig.\ref{fig:CF_T3L1}.
The flat-plate leading edge oscillations of skin-friction coefficient are due to laminar-boundary layer separation.
The results show that the V-SA transition model has difficulty predicting transition onset under these turbulence and flow conditions.
However, transition still occurs later on over the flat-plate. 
The SA model is able to handle these conditions better, determining flow transition at the correct location.

\begin{figure}
	\centering
			\includegraphics[width=0.9\textwidth]{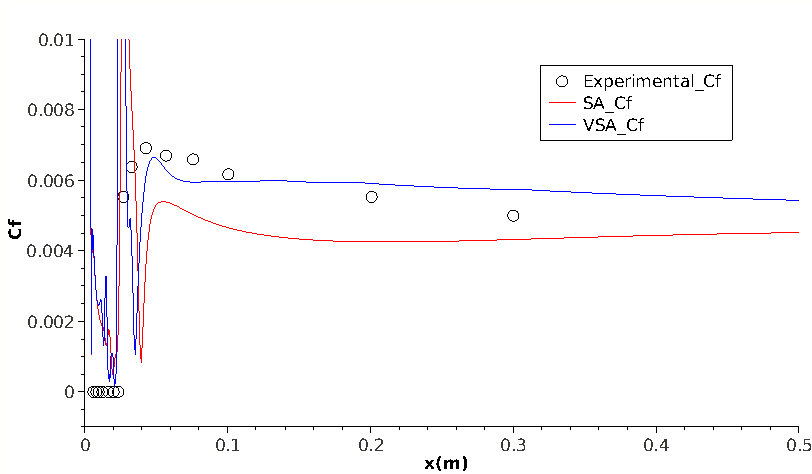}
	
	\caption{Comparison of experimental data from ERCOFTAC T3L3 flat-plate test case skin-friction coefficient distribution with the non-transition SA closure and the transition V-SA model.
The flat-plate leading edge oscillations of skin-friction coefficient are due to flow separation.}
\label{fig:CF_T3L3}
\end{figure}

The second separation induced transition test case considered has a higher turbulence intensity.
The T3L3 ERCOFTAC flat-plate test case results are presented in Fig.\ref{fig:CF_T3L3}.
As can be seen, the V-SA and SA models are able to determine the transition onset very close to the experimental data.
Again the flat-plate leading edge oscillations of skin-friction coefficient are due to boundary layer separation.
It should be noted that both tend to slightly delay transition onset prediction.

\begin{figure}
	\centering
			\includegraphics[width=0.9\textwidth]{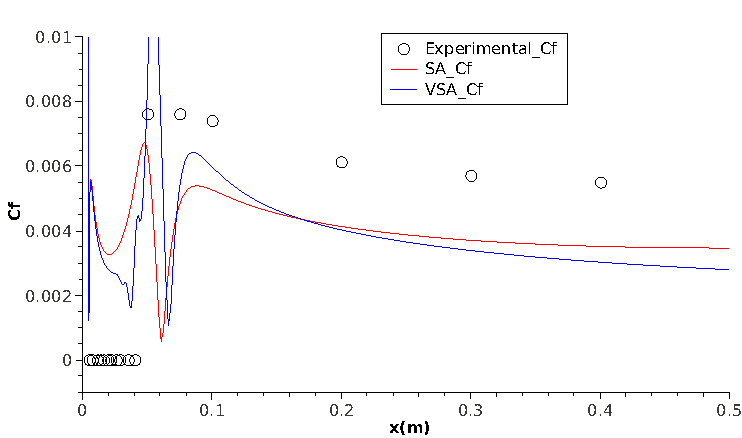}
	
	\caption{Comparison of experimental data from ERCOFTAC T3L5 flat-plate test case skin-friction coefficient distribution with the non-transition SA closure and the transition V-SA model.
The flat-plate leading edge oscillations of skin-friction coefficient are due to flow separation.}
\label{fig:CF_T3L5}
\end{figure}

Finally for the T3L5 test case the results presented in Fig.\ref{fig:CF_T3L5}, indicate that both the V-SA and SA closures are able to determine transition onset near the experimental measurements.
For both cases the flat-plate leading edge oscillations of skin-friction coefficient are due to flow separation.

The incompressible three-dimensional geometry of a 6:1 prolate-spheroid was computed using SA, V-SA and the empirical correlation transition model of Ansys Fluent, the $\gamma-R_{e\theta}$.
These test cases were performed using the experimental data of \cite{Kreplin1985}.
This was obtained through a personal communication with Dr. Kreplin.
The latter experimental work has many test cases, however only some of these were considered for validation purposes.
The selected three test cases had significant transition effects. 
The upstream conditions for these test cases are presented in table \ref{tab:6:1_Prolate_Spheroid_upstream}.

\begin{table}
\caption{6:1 Prolate Spheroid Upstream Conditions}
\label{tab:6:1_Prolate_Spheroid_upstream}
\begin{centering}
{\footnotesize }\begin{tabular}{|c|c|c|c|}
\hline 
{\footnotesize $AoA$} & {\footnotesize $Tu (\%)$} & {\footnotesize $Re$} \tabularnewline
\hline 
{\footnotesize $5\;\mathring{ }$} & {\footnotesize $0.1$} & {\footnotesize $6.5\times10^{6}$} \tabularnewline
\hline
{\footnotesize $15\;\mathring{ }$} & {\footnotesize $0.1$} & {\footnotesize $6.5\times10^{6}$} \tabularnewline
\hline
{\footnotesize $30\;\mathring{ }$} & {\footnotesize $0.1$} & {\footnotesize $6.5\times10^{6}$} \tabularnewline
\hline
\end{tabular}
\par\end{centering}
\end{table}

The experimental setup comprised of a 6:1 prolate-spheroid with a major and minor axis lengths of $2.4$ and $0.4$ meters respectively.
For the considered test cases the experimental flow velocity was $45\left(m/s\right)$.
The experimental fluid kinematic viscosity was $\nu=1.7\times10^{-5}\left(m^{2}/s\right)$.
In order to perform validation using these test cases experimental data, an equal 3D geometry was used with the same dimensions.
The used prolate-spheroid mesh was structured and had $y^+$ values below $0.6$ over the entire surface.
Along the surface major axis the mesh had $400$ computational nodes.
In the azimuth orientation, the prolate-spheroid cross-section had $100$ mesh nodes.
The total value of grid points over the prolate-spheroid surface was $40000$.
The first layer of cells over the spheroid surface were distanced at $1\times10^{-5}$ meters.
The inlet boundary conditions for these test cases of transition under cross-flow effects over a 6:1 prolate-spheroid are presented in table \ref{tab:6:1_Prolate_Spheroid Inlet Boundary Conditions}.
The applied inlet boundary conditions for the Fluent transition model $\gamma-R_{e\theta}$, were selected in order to simulate the same turbulence intensity in the free-stream as presented in table \ref{tab:6:1_Prolate_Spheroid_upstream}. 
\begin{table}
\caption{6:1 Prolate Spheroid Inlet Boundary Conditions}
\label{tab:6:1_Prolate_Spheroid Inlet Boundary Conditions}
\begin{centering}
{\footnotesize }\begin{tabular}{|c|c|c|c|}
\hline 
{\footnotesize $AoA$} & {\footnotesize $k_{p} (m^{2}/s^{2})$} & {\footnotesize $U (m/s)$} & {\footnotesize $\tilde{\nu_{T}}/\nu$} \tabularnewline
\hline 
{\footnotesize $5\;\mathring{ }$} & {\footnotesize $0.003038$} & {\footnotesize $45$} & {\footnotesize $3$} \tabularnewline
\hline
{\footnotesize $15\;\mathring{ }$} & {\footnotesize $0.003038$} & {\footnotesize $45$} & {\footnotesize $3$} \tabularnewline
\hline
{\footnotesize $30\;\mathring{ }$} & {\footnotesize $0.003038$} & {\footnotesize $45$} & {\footnotesize $3$} \tabularnewline
\hline
\end{tabular}
\par\end{centering}
\end{table}
The fluid kinematic viscosity used was the experimental value of $\nu=1.7\times10^{-5}\left(m^{2}/s\right)$.

\begin{figure}
	\centering
			\includegraphics[width=0.9\textwidth]{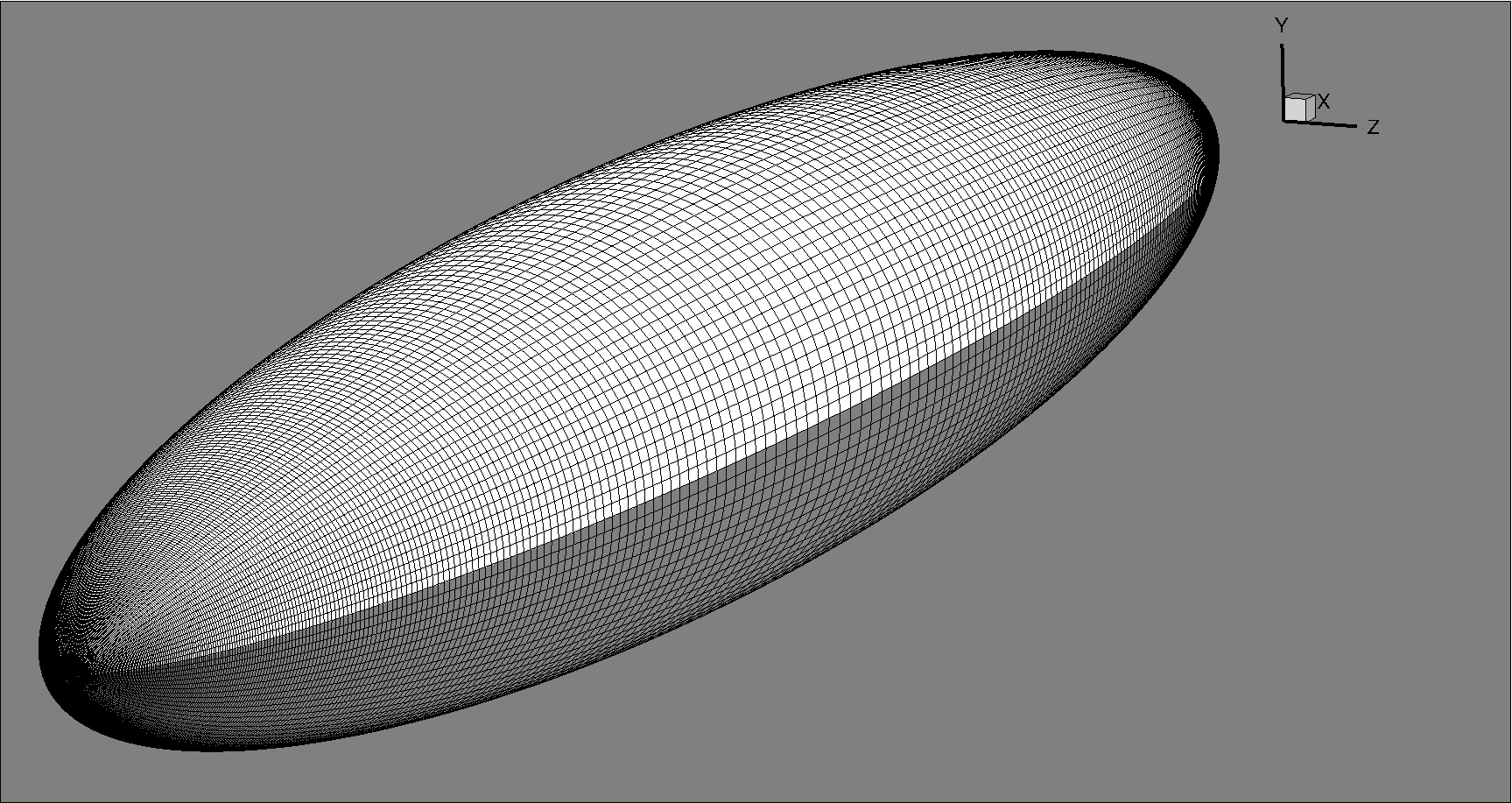}
	
	\caption{X-Z cut-section of 6:1 prolate-spheroid for skin-friction coefficient plots.
The cutting plane is perpendicular to the 6:1 prolate-spheroid minor axis and contains its origin point.
The presented structured mesh has the size of $100\times400$ nodes.}
\label{fig:6:1_Prolate_Spheroid_cut}
\end{figure}

Cut-section plots of skin-friction coefficient for the obtained results is performed.
The cut-section plane is perpendicular to the 6:1 prolate-spheroid minor axis and contains the latter volume center point.
This section cuts the 6:1 prolate-spheroid in the x-z plane as shown in Fig.\ref{fig:6:1_Prolate_Spheroid_cut}.

\begin{figure}
	\centering
			\includegraphics[width=0.9\textwidth]{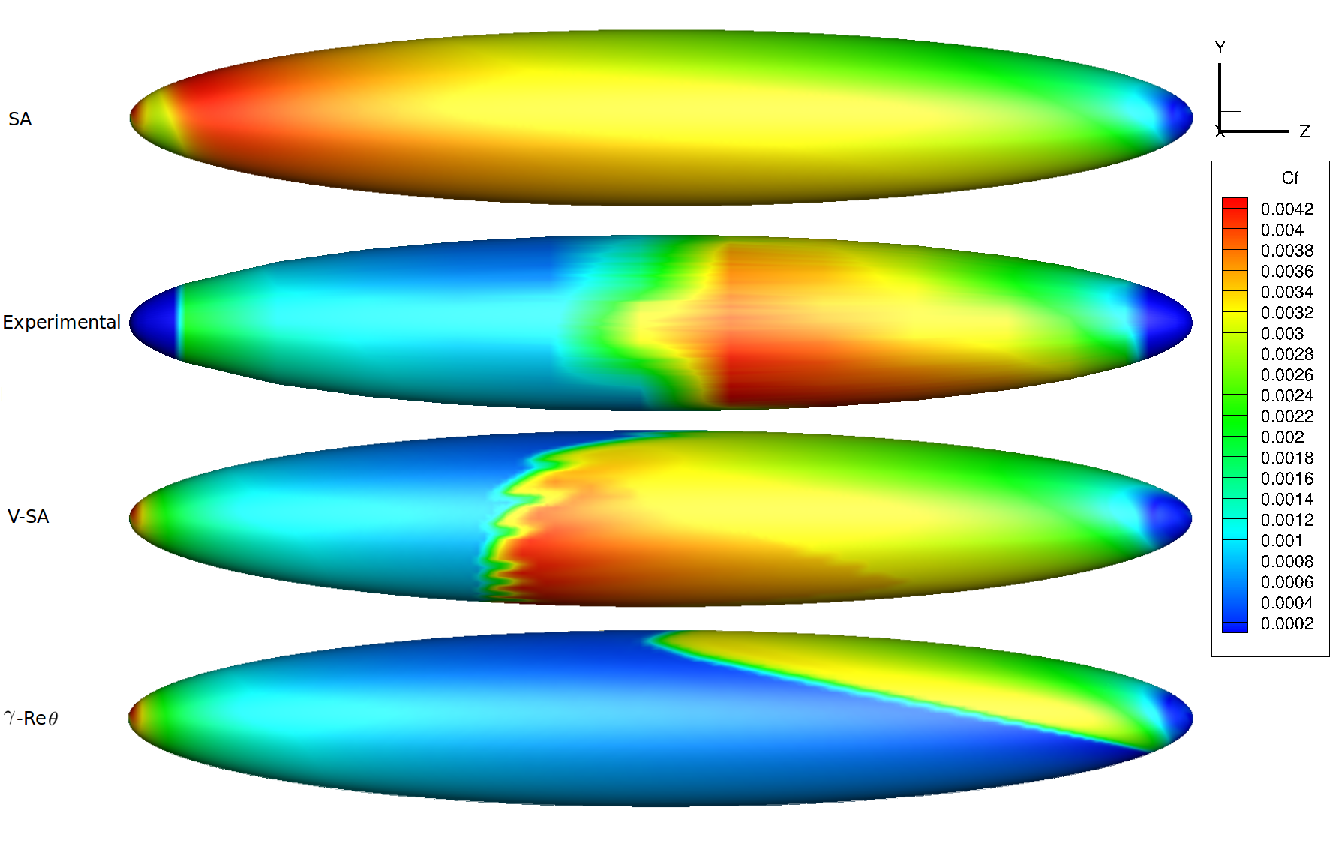}
	
	\caption{Comparison of experimental data of \cite{Kreplin1985} for skin-friction coefficient contour map of 6:1 prolate-spheroid with AoA $5\;\mathring{ }$ with numerical results of non-transitional SA closure and transition V-SA and $\gamma-R_{e\theta}$ models.}
\label{fig:6:1_Prolate_Spheroid_5}
\end{figure}

\begin{figure}
	\centering
			\includegraphics[width=0.9\textwidth]{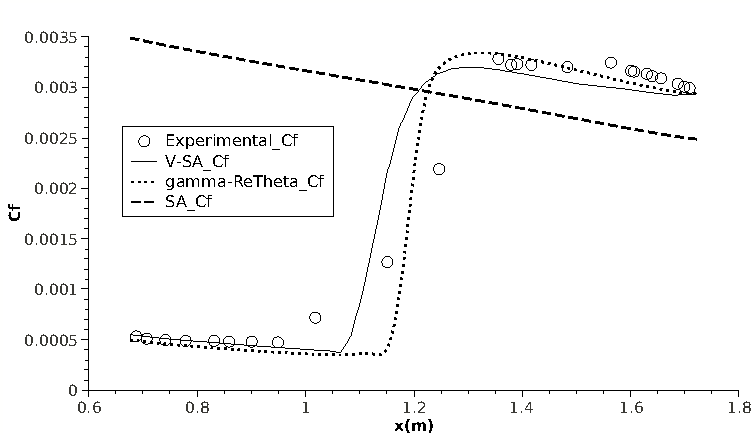}
	
	\caption{Comparison of experimental skin-friction coefficient along a top X-Z cutting plane over the 6:1 prolate-spheroid with AoA $5\;\mathring{ }$ with numerical results of non-transitional SA closure and transition V-SA and $\gamma-R_{e\theta}$ models.}
\label{fig:6:1_Prolate_Spheroid_5_Cf_0_18_slice}
\end{figure}

\begin{figure}
	\centering
			\includegraphics[width=0.9\textwidth]{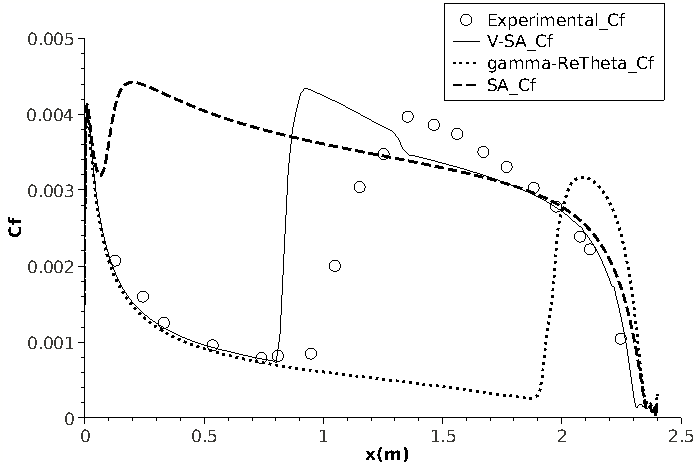}
	
	\caption{Comparison of experimental skin-friction coefficient along the X-Z cutting plane over the 6:1 prolate-spheroid with AoA $5\;\mathring{ }$ with numerical results of non-transitional SA closure and transition V-SA and $\gamma-R_{e\theta}$ models.}
\label{fig:6:1_Prolate_Spheroid_5_Cf}
\end{figure}

The results for the $5\;\mathring{ }$ angle of attack, AoA, test case are presented in Fig.\ref{fig:6:1_Prolate_Spheroid_5}.
The experimental data for the prolate-spheroid tips is not available.
Thus these have been assigned with a zero value of skin-friction coefficient in all presented results.
However, it must be noted that for all AoA, the flow separates at the trailing edge of the prolate-spheroid.
As can be seen, the SA closure determines transition onset right at the leading edge of the prolate-spheroid.
The V-SA transition model predicts transition onset near to the experimental result.
It should be noted that, the transition length of the V-SA is shorter than the experimental data.
However the transition onset position is quite close to the experimental measurements.
The $\gamma-R_{e\theta}$ transition model predicts the transition onset correctly but the transition line along the surface has an incorrect angle.
A skin-friction coefficient plot from a top x-z cutting plane parallel to the one presented in Fig.\ref{fig:6:1_Prolate_Spheroid_cut} is presented in Fig.\ref{fig:6:1_Prolate_Spheroid_5_Cf_0_18_slice}.
It is shown that the $\gamma-R_{e\theta}$ empirical transition model is able to correctly predict transition onset as well as the late transition value of skin-friction coefficient.  
The central x-z cutting plane results are presented in Fig.\ref{fig:6:1_Prolate_Spheroid_5_Cf}.
The presented results confirm the reliability of the V-SA model.
  
\begin{figure}
	\centering
			\includegraphics[width=0.9\textwidth]{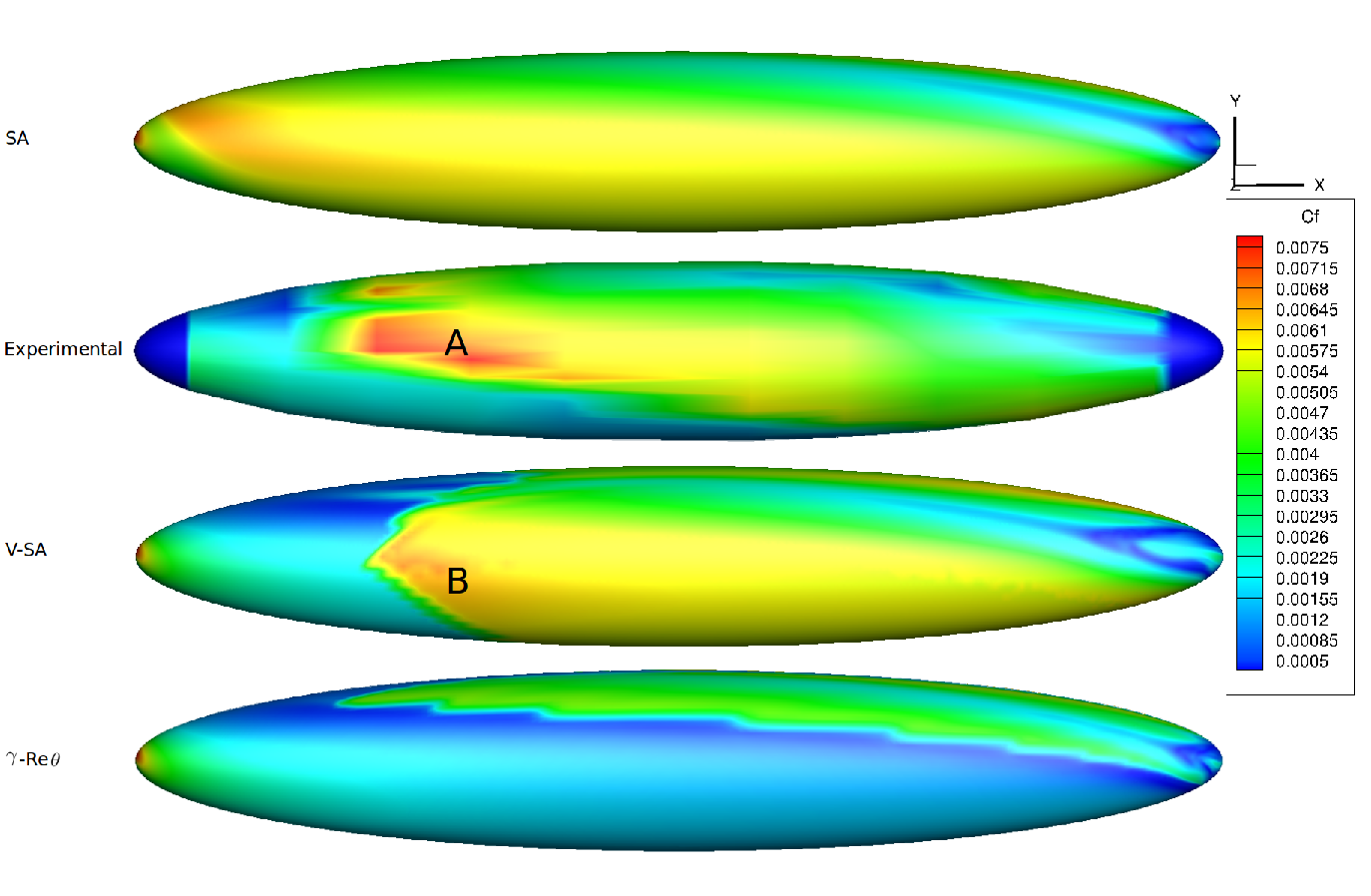}
	
	\caption{Comparison of experimental skin-friction coefficient contour map of 6:1 prolate-spheroid with AoA $15\;\mathring{ }$ with numerical results of non-transitional SA closure and transition V-SA and $\gamma-R_{e\theta}$ models.
In the regions marked by the letters A and B there is a severe cross-flow transition effect.}
\label{fig:6:1_Prolate_Spheroid_15}
\end{figure}

\begin{figure}
	\centering
			\includegraphics[width=0.9\textwidth]{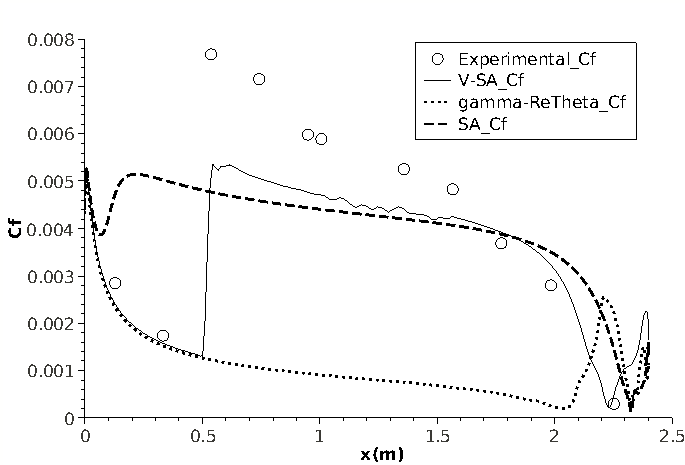}
	
	\caption{Comparison of experimental skin-friction coefficient along the X-Z cutting plane over the 6:1 prolate-spheroid with AoA $15\;\mathring{ }$ with numerical results of non-transitional SA closure and transition V-SA and $\gamma-R_{e\theta}$ models.}
\label{fig:6:1_Prolate_Spheroid_15_Cf}
\end{figure}

\begin{figure}
	\centering
			\includegraphics[width=0.9\textwidth]{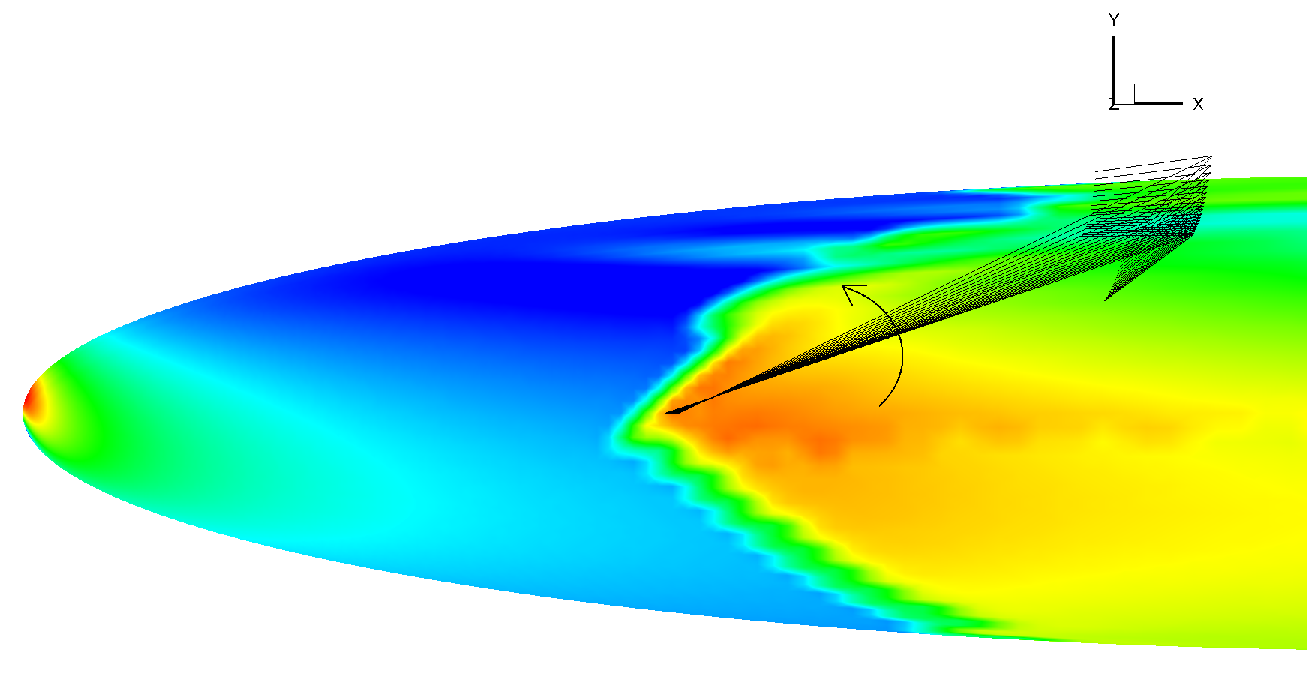}
	
	\caption{Cross-flow effect on velocity profile normal to the surface of the 6:1 prolate-spheroid with AoA $15\;\mathring{ }$.
The V-SA transition model calculated skin-friction coefficient contour map is used as surface contour.}
\label{fig:6:1_Prolate_Spheroid_15_Crossflow}
\end{figure}

\begin{figure}
	\centering
			\includegraphics[width=0.9\textwidth]{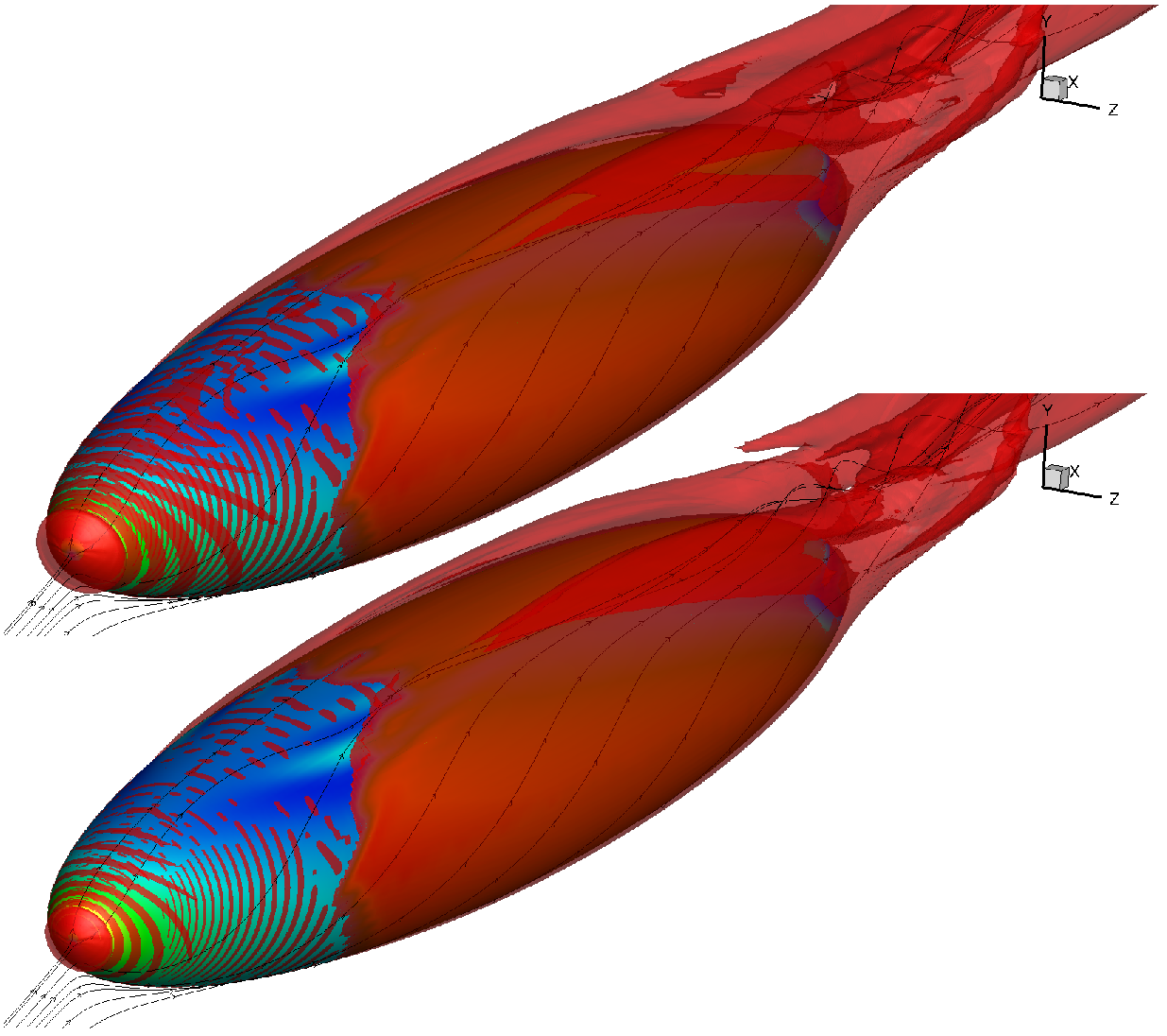}
	
	\caption{Side view of red colored transparent $\overline{u^{'}v^{'}}$ iso-surfaces with flow streamlines over the 6:1 prolate-spheroid with AoA $15\;\mathring{ }$.
Top image represents the $\overline{u^{'}v^{'}}$ iso-surface equal to -0.02.
Bottom image represents the $\overline{u^{'}v^{'}}$ iso-surface equal to -0.03.
The V-SA transition model calculated skin-friction coefficient contour map is used as surface contour.}
\label{fig:6:1_Prolate_Spheroid_15_uv_side}
\end{figure}

\begin{figure}
	\centering
			\includegraphics[width=0.9\textwidth]{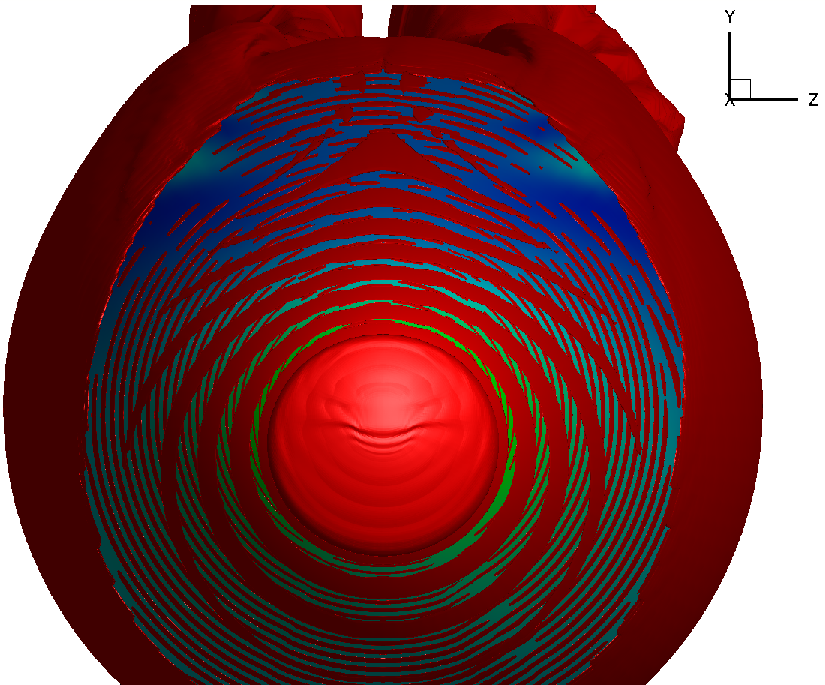}
	
	\caption{Front view of red colored non-transparent $\overline{u^{'}v^{'}}$ iso-surfaces over the 6:1 prolate-spheroid with AoA $15\;\mathring{ }$.
The image represents the $\overline{u^{'}v^{'}}$ iso-surface equal to -0.03.
The V-SA transition model calculated skin-friction coefficient contour map is used as surface contour.}
\label{fig:6:1_Prolate_Spheroid_15_uv_front}
\end{figure}

The results for the AoA $15\;\mathring{ }$ test case shown in Fig.\ref{fig:6:1_Prolate_Spheroid_15} expose the fact that the SA model predicts transition at the leading edge of the geometry.
The V-SA closure is able to predict transition onset near the experimental values although with some delay.
Again the $\gamma-R_{e\theta}$ transition model predicts the transition onset very close to the experimental data but the transition zone shape is incorrect. 
In the work of \cite{Seyfert2012}, similar numerical results were obtained with the $\gamma-R_{e\theta}$ transition model for these last two test cases.
The x-z cutting plane results are presented in Fig.\ref{fig:6:1_Prolate_Spheroid_15_Cf}.
As can be seen, the V-SA model predicts transition onset close to the experimental data.
However, the predicted transition skin-friction coefficient peak value is lower than that of the experimental data.
Also, for this test case a velocity profile normal to the prolate-spheroid surface was taken in order to observe the cross-flow effect.
The velocity profile twist due to the present cross-flow conditions is shown in Fig.\ref{fig:6:1_Prolate_Spheroid_15_Crossflow}.
The evolution of flow streamlines and $\overline{u^{'}v^{'}}$ iso-surfaces over the spheroid are presented in Fig.\ref{fig:6:1_Prolate_Spheroid_15_uv_side}.
As can be observed, there is flow separation at the trailing edge of the spheroid.
Also, the leading edge $\overline{u^{'}v^{'}}$ iso-surfaces patterns are quite interesting.
There seems to be two sets of $\overline{u^{'}v^{'}}$ fluctuations over the spheroid nose.
This can also be observed in the front view of the 6:1 prolate-spheroid $\overline{u^{'}v^{'}}$ iso-surface patterns.
These are presented in Fig.\ref{fig:6:1_Prolate_Spheroid_15_uv_front}.
In the fully turbulent flow region these patterns cease to exist.
Instead a constant iso-surface covers the remaining extension of the prolate-spheroid.

\begin{figure}
	\centering
			\includegraphics[width=0.9\textwidth]{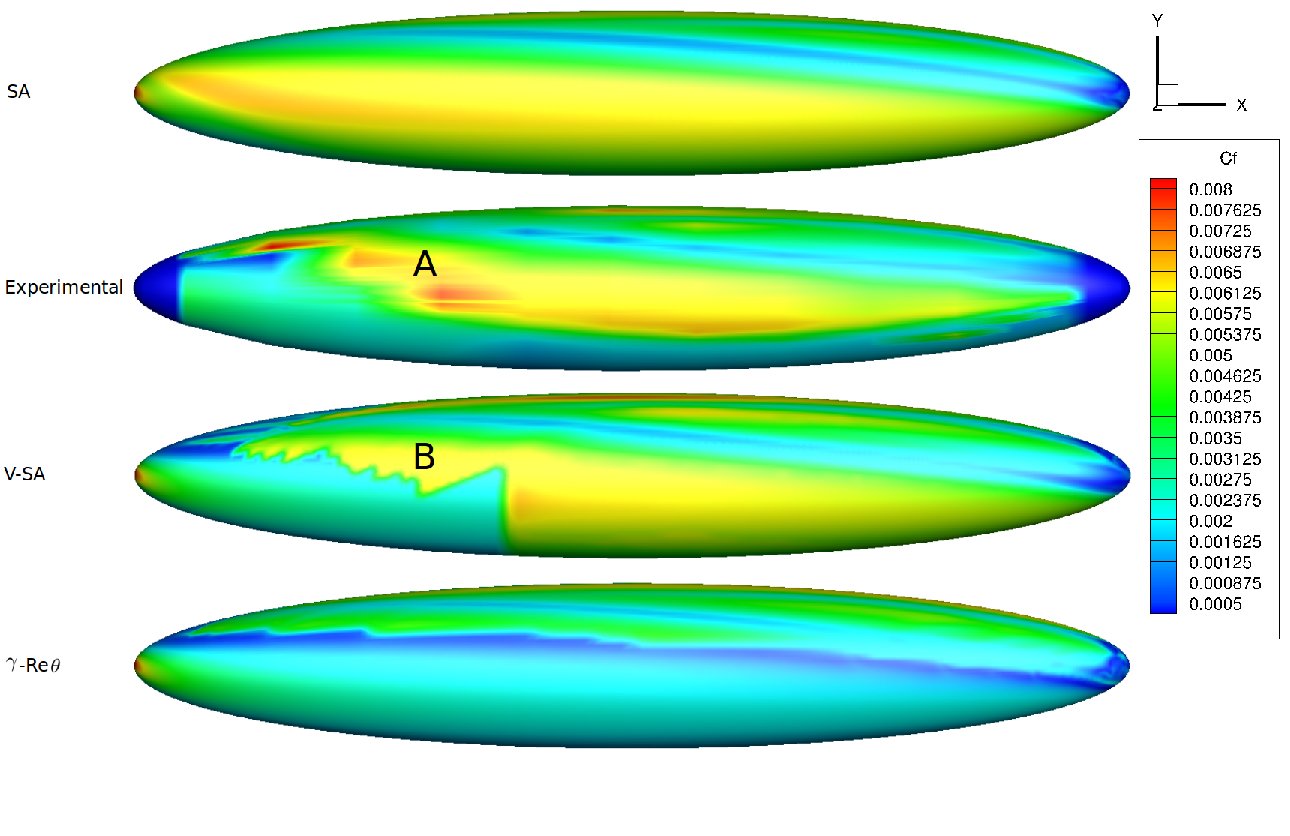}
	
	\caption{Comparison of experimental skin-friction coefficient contour map of 6:1 prolate-spheroid with AoA $30\;\mathring{ }$ with numerical results of non-transitional SA closure and transition V-SA and $\gamma-R_{e\theta}$ models.
In the regions marked by the letters A and B there is a severe cross-flow transition effect.}
\label{fig:6:1_Prolate_Spheroid_30}
\end{figure}

\begin{figure}
	\centering
			\includegraphics[width=0.9\textwidth]{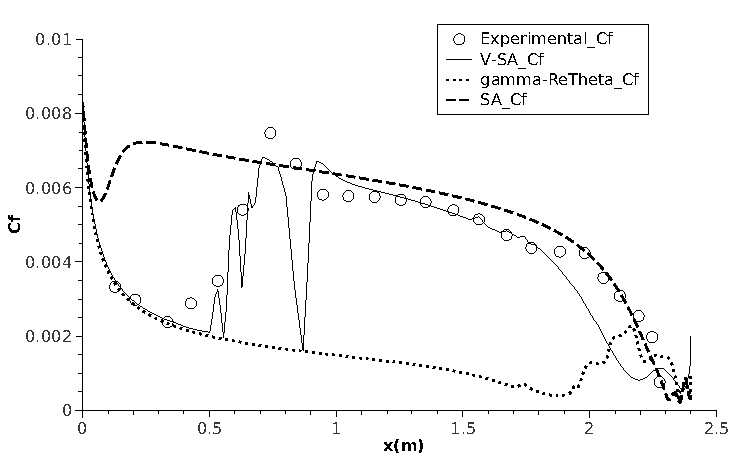}
	
	\caption{Comparison of experimental skin-friction coefficient along the X-Z cutting plane over the 6:1 prolate-spheroid with AoA $30\;\mathring{ }$ with numerical results of non-transitional SA closure and transition V-SA and $\gamma-R_{e\theta}$ models.}
\label{fig:6:1_Prolate_Spheroid_30_Cf}
\end{figure}

The final 6:1 prolate-spheroid validation test case was performed with an AoA of $30\;\mathring{ }$.
As can be seen in the results of Fig.\ref{fig:6:1_Prolate_Spheroid_30}, the V-SA model's transition onset prediction is in accordance with the experimental measurements.
The V-SA transition closure can even predict the saw-tooth shape behavior of the transition onset line very similar to the experimental data.
The SA turbulence model predicts transition onset at the beginning of the prolate-spheroid.
The $\gamma-R_{e\theta}$ exhibits a similar behavior to the last two test cases presented here.
The x-z cutting plane results are presented in Fig.\ref{fig:6:1_Prolate_Spheroid_30_Cf}.
As shown the V-SA model is able to predict transition onset near the experimental data and calculates the transition process with a saw-shape.
The latter shape is due to strong cross-flow effects during turbulence transition.
This effect is visible in the regions marked by the letters A and B in Fig.\ref{fig:6:1_Prolate_Spheroid_30}. 

\section{Conclusions}
\label{sec:11}
The rational of the transition V-model development was presented.
A detailed description of the mechanical approximation model was performed.
The transition model pre-transitional turbulent kinetic energy transport equation formulation was derived.

The model was validated for eight benchmark test cases.
The V-SA transition model was validated in the zero-pressure gradient flat-plate T3A test case.
The V-model predicted values of $\overline{u^{'}v^{'}}$ in the pre-transition boundary layer were validated with experimental data from the flat-plate T3A ERCOFTAC database.
As stated by \cite{Bradshaw1994}, ''the so-called inactive motion... is simple: the motion near the surface, ... results mainly from eddies actually generated near the surface, ... the contribution... to the shear stress -$\rho\overline{u^{'}v^{'}}$ is small``.
The transition V-model predicts this small contribution.
Also the V-SA model was validated with the flat-plate pressure-gradient ERCOFTAC T3C3 test case.
The experimental ERCOFTAC T3L flat-plate test cases were used to further validate the V-SA model.
These were used for validation of the V-model under separation induced transition.
For the very low turbulence intensity case of Tu=$0.2\%$, T3L1, the model predicted a delayed transition onset.
This is possibly related to an excessive effect of the V-model destruction term.
Since the free-stream turbulence intensity is very low, with low free-stream velocity, so is its turbulent kinetic energy.
Therefore there seems to be an excessive sensibility to the destruction effect by the separation bubble vorticity.
For all of the remaining T3L tested cases the model behaved correctly.

The strengths of the V-SA model are verified on the transition under cross-flow effects test cases of the three-dimensional 6:1 prolate-spheroid geometry.
It was observed that although the free-stream turbulence intensity for these cases is very low, Tu=$0.1\%$, the model is able to predict transition onset patterns close to the experimental data.
However, the transition length of the model is short in one of the tested cases.
For the test case with a low angle of attack of $5\;\mathring{ }$, the V-SA model predicts a short transition length.
Although the reason for the latter is unclear, it is suspected that an excessive pre-transition turbulent kinetic energy diffusion inside the boundary layer might be the reason for such short transition length.
The rate of turbulence intermittency diffusion into the transition boundary layer has a major role determining transition length as shown in the work of \cite{Durbin20121}. 

For further research the transition length control of the model should be investigated as well as the delayed transition onset under the combination of very low free-stream turbulence intensity and low speed conditions.  

\begin{acknowledgements}
The current work was performed as part of Project MAAT, supported by European Union on course of the 7th Framework Programme, under grant number 285602 and was also supported by
C-MAST, Center for Mechanical and Aerospace Science and Technologies research unit No.151.
\end{acknowledgements}

\bibliographystyle{abbrvnat}      
\bibliography{base_dados_transicao_30_Jan_2014}


\end{document}